 \documentclass[11pt,twoside,reqno,centertags]{amsart}
 

 \usepackage{amsmath,amsthm,amsfonts,amssymb}
\usepackage[mathscr]{euscript}
\usepackage[hidelinks]{hyperref} 
 \usepackage{mathrsfs} 
\usepackage{graphicx}
\usepackage{color}
  \usepackage{epsfig}

\usepackage{graphics}
\usepackage{mathtools}
\usepackage{enumerate}
\usepackage{cite}
\usepackage{amsaddr}

\usepackage{float}
\usepackage{comment}

\newcommand\EN           {\end{equation}}
\newcommand\bes           {\begin{subequations}}
\newcommand\esu           {\end{subequations}}

\newcommand\p            {\partial}

\newcommand{\ep}{\mathbf{e}_{\varphi}}
\newcommand{\et}{\mathbf{e}_{\theta}}
\newcommand{\er}{\mathbf{e}_r}

\newcommand{\bu}{\boldsymbol{u}}
\newcommand{\x}{\mathsf{x}}
\newcommand{\y}{\mathsf{y}}
\newcommand{\z}{\mathsf{z}}
\newcommand{\csf}{\mathsf{c}}

\newcommand{\ea}{\mathbf{e}_{1}}
\newcommand{\eb}{\mathbf{e}_{2}}
\newcommand{\ec}{\mathbf{e}_{3}}
\newcommand{\td}{\theta^{\dag}}
\newcommand{\pd}{\varphi^{\dag}}

\renewcommand{\H}{\mathscr{H}}
\newcommand{\A}{\mathscr{M}}

\begin{document}
 
\title{ Near-inertial Pollard waves \\
 modeling the arctic halocline }
\thanks{Accepted for publication in \emph{Differential and Integral Equations}} 
\date{}
\maketitle     
 
\vspace{ -1\baselineskip}

{\small
\begin{center}
 {\sc Christian Puntini} \\
Faculty of Mathematics, University of Vienna\\
 Oskar-Morgenstern-Platz 1, 1090\\
Vienna, Austria \\christian.puntini@univie.ac.at
\end{center}
}

\numberwithin{equation}{section}
\allowdisplaybreaks

\smallskip

 \begin{quote}
\footnotesize
{\bf Abstract.}  
  We present an explicit and exact solution to the governing equations describing the vertical structure of the Arctic Ocean region centred around the North Pole. The solution describes a stratified water column with three constant-density regions: a motionless bottom layer, a middle layer---the halocline---described by nonhydrostatic, near-inertial Pollard waves, and an upper layer presenting a mean current and a wave motion associated with the one in the halocline layer.
\end{quote}

\section{Introduction}
\begin{figure}
    \centering
    \includegraphics[width=.75\linewidth]{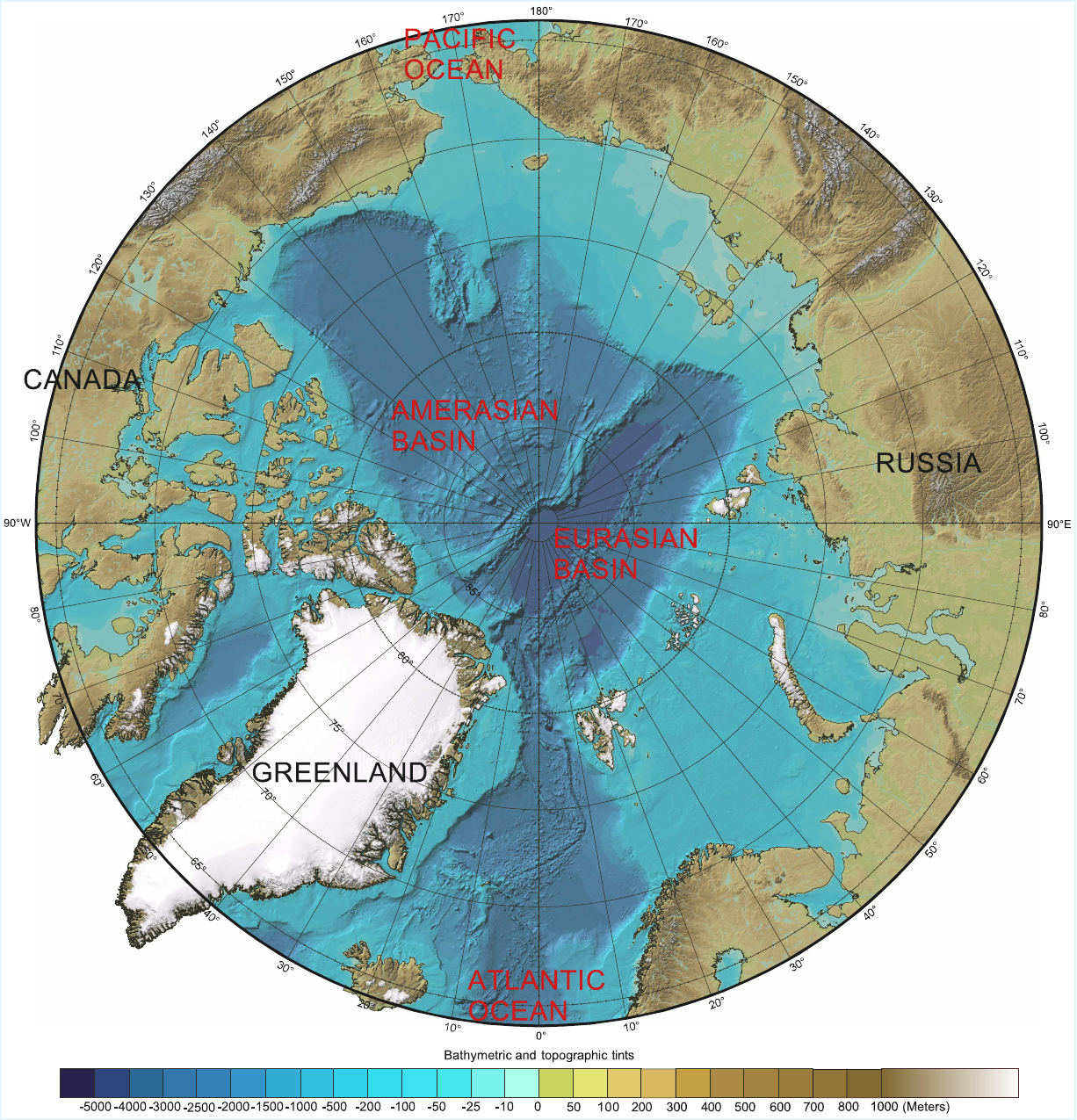}
    \caption{Map of the Arctic Ocean. Adapted from the IBCAO map \texttt{https://www.ngdc.noaa.gov/mgg/bathymetry/arctic/currentmap.html}. Credit \cite{IBCAO}.}
    \label{fig-ARCTIC}
\end{figure}

\begin{figure}[ht]
\centering
\includegraphics[width=\linewidth]{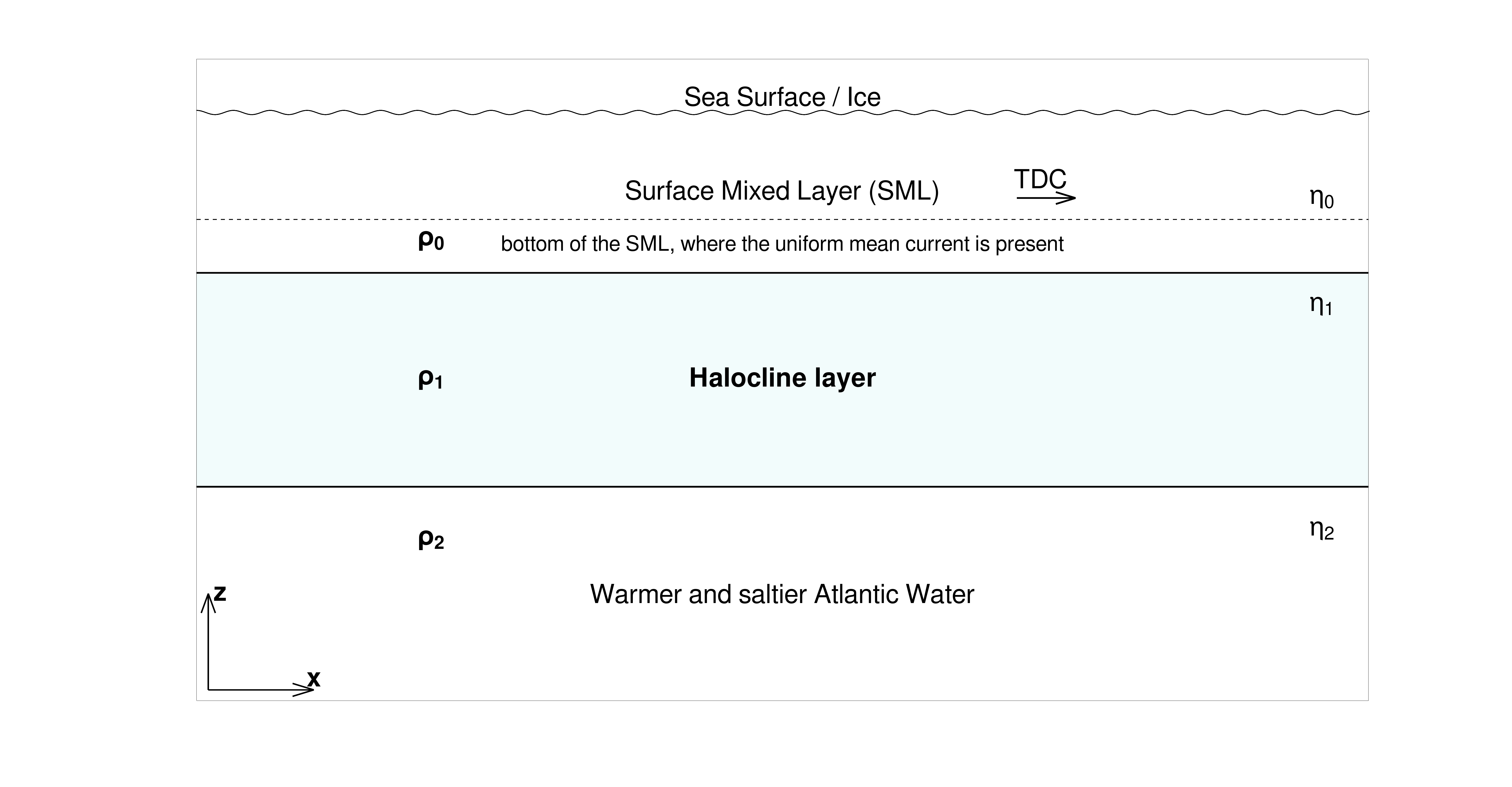}
\caption{A depiction of the model we are considering.}
\label{strati}
\end{figure}

The Arctic Ocean, located in the Northern Hemisphere and encompassing the North Pole, is a small ocean about $4000\, \mathrm{km}$ long and $2500\, \mathrm{km}$ wide, covering an area of around 10 million $\, \mathrm{km}^2$. It includes two main deep basins (around $4000\, \mathrm{m}$ deep) separated by the underwater Lomonosov Ridge, the Amerasian Basin (divided into the Makarov and Canada basins), and the smaller Eurasian Basin (divided into the Amundsen and Nansen basins), and is surrounded by shallow seas (less than $400\, \mathrm{m}$ deep): the Barents Sea, the Kara Sea, the Laptev Sea, the Siberian Sea, the Chukchi Sea and the Beaufort Sea (see \cite{BertosioPHD} or \cite{Talley}). See Figure \ref{fig-ARCTIC}.\\
The primary oceanic inflows are from the Atlantic Ocean through the Fram Strait and the Barents Sea, and from the Pacific Ocean via the Bering Strait. Significant freshwater inflows also come from rivers in North America and Siberia. These inflows are major factors influencing the salinity and temperature variability of the water column, with salinity playing a key role in stratification, creating a halocline rather than a thermocline (which is instead typical of mid-latitudes and equatorial regions). As a matter of fact, one of the main features of the Arctic Ocean is that it is predominantly stratified by salinity (hence it is defined as a \(\beta\)-ocean) rather than by temperature (the so called \(\alpha\)-oceans) (see\cite{BertosioPHD} or \cite{Carmack2007}).\\
Moreover, the central region of Arctic Ocean---centred around the North Pole and subject of our analysis---is entirely covered by a relatively thin (not exceeding $2\, \mathrm{m}$ of thickness) layer of sea ice, which reduces in summer when temperature increases. Due to the global warming, the sea ice is rapidly declining, and the Arctic region is warming faster than the global average, due to the process known as Arctic amplification. This makes the Arctic highly vulnerable to climate change, potentially leading to an ice-free Arctic Ocean in summer and with thinner and more mobile sea ice in winter. The transition to a seasonally ice-free Arctic will profoundly affect Arctic oceanography, the marine ecosystems it supports, and the global climate (see \cite{TM2020}).\\
In addition to the atmospheric conditions, the Arctic sea ice is influenced by the sea water beneath it. The Arctic Ocean is stratified into a cold and fresh surface mixed layer (SML), with a depth between $5$ to $100$ meters, a halocline below the mixed layer with a base depth ranging from $40$ to $200$ meters, and a layer of warmer and saltier Atlantic Water (AW) (see \cite{MetznerSalzmann}). The depths of the boundaries of this layers varies mainly according to temperature and presence of ice. In general, it can be said that in winter, with lower temperatures and increased ice presence, their depth increases, whereas in summer, these decrease (see \cite{Peralta}).\\
The halocline is a region of strong stratification, which prevents interaction of the ice cover with AW heat by the direct surface-generated mixing of the SML (see \cite{Polyakov2018} and \cite{Guthrie} for an in-detail description of the Arctic water column and its main physical processes); consequently, the halocline layer, situated above the saltier and denser water, is fundamental for the formation of the ice cover (see \cite{Polyakov2018}). \\
The physical processes in the mixed layer, mainly the ice-motion induced by wind and the Transpolar Drift Current (TDC)---the Arctic Ocean current transporting surface waters and sea ice from the Laptev Sea and the East Siberian Sea towards Fram Strait, with a speed of around $0.07\, \mathrm{ms^{-1}}$ around the North Pole---induces a current with magnitude approximately $0.1\, \mathrm{ms^{-1}}$ on the lower boundary of the mixed layer (see \cite{Guthrie} and data therein). \\
The presence of the permanent ice layer makes physical measurements particularly difficult. Hence, due to the lack of data from observation, the theoretical analysis of this region is particularly useful.\\
The aim of this paper is to describe the halocline layer via an explicit and exact solution of the (approximated) nonlinear equations governing the ocean dynamics. Such a solution will represent nonlinear waves propagating parallel to the Transpolar Drift Current. Just above the halocline, we consider a mean current coupled with a nonlinear wave motion, and the layer of Atlantic Water under the halocline is considered in a hydrostatic state.\\
The solution in this study is constructed by adapting Pollard's solution, presented in \cite{Pollard} for surface waves accounting for the effects of Earth's rotation and extending the remarkable solution provided by Gerstner in \cite{Gerstner}. See also \cite{C2001}, \cite{ConstantinBook} and \cite{Henry}, and \cite{CW} for more general trochoidal solutions. Recently, Gerstner-like solutions were used to describe equatorial waves \cite{Constantin2012}, \cite{Constantin2013}, \cite{Constantin2014}, \cite{ConstatinEquatorial}, \cite{Henry16} and to study their linear instability \cite{CG}, \cite{HStability}, \cite{HC15},  or to study wave-current interactions \cite{Constantin2017}, \cite{K17} \cite{McCarney2023}, \cite{McCarney2024} and edge waves \cite{C  Edge waves}, \cite{IK2015} \cite{Weber}, \cite{Miao}. See also \cite{Hsu14}, \cite{R-S2018}, \cite{Hsu17} and \cite{ChuETAL} for Gerstner-like waves in other oceanographic models.\\
Similarly, Pollard-like internal wave approaches have been developed in \cite{IK15} or \cite{K19}, and their instability has also been subject of investigation \cite{IK shortw}, \cite{IK stability}.
 \\
Despite the fact that within the halocline one could distinguish between the cold halocline layer in the Eurasian Basin, the Pacific Halocline Waters in the Amerasian Basin and the lower halocline water (see \cite{MetznerSalzmann}), the simplified model we will consider features constant densities in the layers under consideration. The case of depth-dependent densities could be considered in future works, even if density variations are very small within each stratum (see \cite{Polyakov2018}), so accounting for density variation would probably not provide substantial differences. Our model features three constant densities, $\rho_0$, $\rho_1$ and  $\rho_2$, with $\rho_0<\rho_1<\rho_2$, where $\rho_0$ represents the fresh and cold water of the surface mixed layer above the halocline, $\rho_1$ is the density of the halocline's water and $\rho_2$ is the density of the saltier and warmer water (AW) below the halocline. See Figure \ref{strati}. \\
\\
More precisely, we assume the following water characteristics (we refer to \cite{Rudels}): 
\begin{itemize}
    \item Surface Mixed Layer: water density $\rho_0$, $\mathfrak{T}=-1.5^{\circ}\, \mathrm{C}$, $\mathfrak{S}=34.0\, \mathrm{psu}$;
       \item Halocline: water density $\rho_1$, $\mathfrak{T}=0^{\circ}\, \mathrm{C}$, $\mathfrak{S}=34.2\, \mathrm{psu}$;
       \item Atlantic Water: water density $\rho_2$, $\mathfrak{T}=2^{\circ}\, \mathrm{C}$, $\mathfrak{S}=34.9\, \mathrm{psu}$;
\end{itemize}
and density variations can be modeled by (see \cite{BertosioPHD} or \cite{Talley})
\begin{equation}\label{densityvariation}
    \frac{d\rho}{\rho}=-\alpha\, d\mathfrak{T}+\beta\, d\mathfrak{S},
\end{equation}
where $\mathfrak{T}$ is the potential temperature, $\mathfrak{S}$ is the salinity, $\alpha\approx53\cdot10^{-6}\,\mathrm{K^{-1}}$ is the thermal expansion coefficient and $\beta\approx785\cdot10^{-6}\,\mathrm{kg}\,\mathrm{g^{-1}}$ is the haline contraction coefficient. Both $\alpha$ and $\beta$ depend of the water properties, therefore the value reported here refer only for the Arctic Ocean (see \cite{Talley}).\\
\\
\noindent
The paper is structured as follow:
\begin{itemize}
    \item in Section 2 we derive the governing equations. More precisely, as we are investigating the fluid dynamics around the North Pole, where the classical spherical coordinates fail, we need to adopt the rotated spherical coordinates developed in \cite{CJ2023}, and therefore adapt the tangent-plane, traditional and $f$-plane approximations to this new coordinate system, starting from the Euler equations. Then, by a $2D$ rotation we will align one of the axes with the Transpolar Drift Current in order to simplify the study of the internal waves propagating in its direction;
    \item Section 3 is devoted to the main result of our analysis. We find an explicit solution, using the Lagrangian description of the flow, detailing the internal wave motion in the halocline by near-inertial Pollard waves propagating parallel to the Transpolar Drift Current. The dispersion relation for the nonlinear waves is obtained by imposing the dynamic boundary condition (namely, the continuity of the pressure across the two layers: the top and the bottom of the halocline).\\
    In previous works \cite{Constantin2017}, \cite{Kluczek}, \cite{McCarney2024}, with only one dynamic boundary condition, two modes of the wave motion were found, one fast mode standard to the theory of internal waves, and a second one, slow, with the period close to the inertial period of the Earth $T_i=\frac{2\pi}{f}$. Constantin \& Monismith \cite{Constantin2017} refer to this slow-mode wave as inertial Gerstner wave.\\
    In this work, imposing two dynamic conditions, one per boundary of the halocline, only the slow mode is shown to be relevant. Henceforth, the waves will be near-inertial;
   \item Section 4 is aimed at the review of some properties of the flow: vorticity, Lagrangian and Eulerian mean velocities, Stokes drift and mass flux;
    \item in Section 5 we study the linearised version of the problem, still adopting the Lagrangian approach. It is shown that, even if the solution found in the nonlinear analysis is a solution also for the linearised equations of motion (in the Lagrangian setting), the dynamic boundary condition cannot be satisfied, highlighting the fact that the nonlinearity of the equations plays a key role in the model; 
    \item finally, we conclude in Section 6 with a discussion on the results.
\end{itemize}
\section{Governing (approximated) equations}
\subsection{Classical and rotated spherical coordinates}
\begin{figure}[ht!]
						\centering
							\includegraphics[width=.48\linewidth]{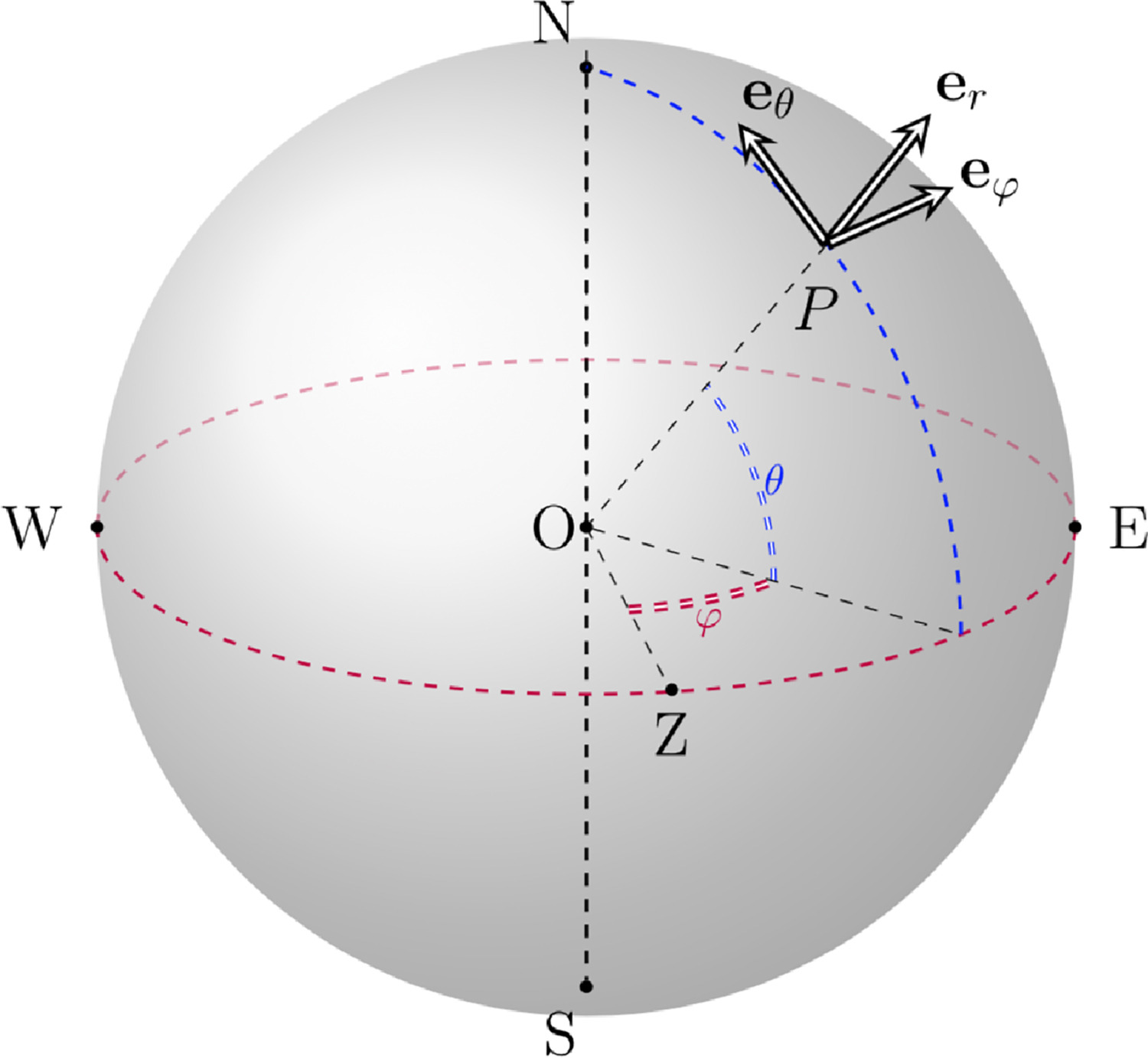}
						\includegraphics[width=.48\linewidth]{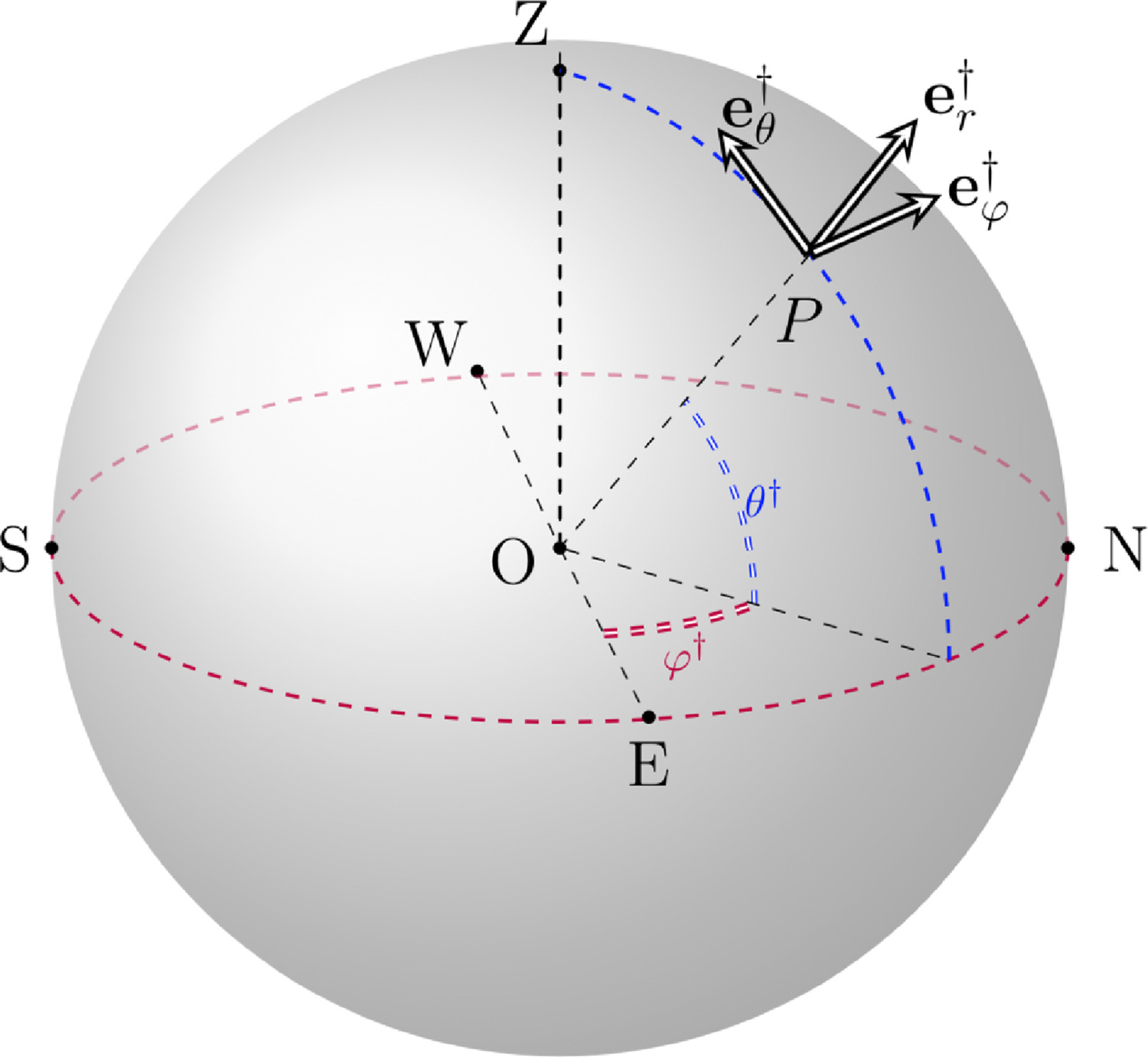}
							
					\caption{Left: classical spherical coordinates. Right: rotated spherical coordinate. Images from \cite{CJ2023}. CC BY 4.0 (\texttt{http://creativecommons.org/licenses/by/4.0}).}
				\label{coordinates}
\end{figure}

We consider the Earth to be a sphere of radius $R\approx6371\, \mathrm{km}$. The classical spherical coordinate system is not suitable for the study of flows in regions centred around the North (and South) Pole, as longitude is undefined at the poles. This result is a consequence of the `hairy ball theorem' (see \cite{Chinn} and \cite{CJBook}).
In this section, we review the construction of the rotated spherical coordinate system developed in \cite{CJ2023} (see also \cite{CJBook} for a detailed exposition) to avoid this problem, and we derive the $f$-plane approximation for the Euler equations in this new coordinate system.\\
Let us start by considering the standard Cartesian coordinate system $(X,Y,Z)$ with basis $(\mathbf{e}_1, \mathbf{e}_2, \mathbf{e}_3) $ positioned at the center $O$ of the Earth, and pointing in the direction $\Vec{OZ}$ of the Null Island,  $\Vec{OE}$ East,  $\Vec{ON}$ North, respectively, and define the classical spherical coordinates $(\varphi,\theta, r)$, with $\varphi \in [0,2\pi)$ and $\theta\in [-\frac{\pi}{2},\frac{\pi}{2}]$ being the angles of longitude and latitude respectively, and $r$ the distance from Earth's center. With respect to these coordinates, a point $P$ has position vector

\begin{equation}\label{r}
	\Vec{OP}:=\mathbf{r}=r\cos\theta\cos\varphi \ea+r\cos\theta\sin\varphi \eb+r\sin\theta \ec.
\end{equation}
Namely, the change of coordinates is given by
\begin{equation}\label{rot coord}
	\left\{ \begin{array}{ll}
		X=r \cos\theta \cos \varphi,\\
		Y=r \cos\theta \sin \varphi,\\
		Z=r \sin\theta ,
	\end{array}\right.
	\qquad
	\left\{ \begin{array}{ll}
		\varphi = \tan^{-1}\left(\frac{Y}{X}\right),\\
		\theta = \sin^{-1}\left(\frac{Z}{\sqrt{X^2+Y^2+Z^2}}\right),\\
		r=\sqrt{X^2+Y^2+Z^2}.
	\end{array}\right.
\end{equation}
Now, let us define a new Cartesian coordinate system  $(\mathbf{e}_1^{\dag}, \mathbf{e}_2^{\dag}, \mathbf{e}_3^{\dag}) $ by permuting cyclically the first three Cartesian axes
\begin{equation}\label{4.3}
	\mathbf{e}_1^{\dag}=\ec,\quad \mathbf{e}_2^{\dag}=\ea, \quad\mathbf{e}_3^{\dag}=\eb.
\end{equation}
The associated Cartesian coordinates to $(\mathbf{e}_1^{\dag}, \mathbf{e}_2^{\dag}, \mathbf{e}_3^{\dag}) $ are $(X^{\dag},Y^{\dag},Z^{\dag})$, and, in terms of the associated azimuthal $\theta^{\dag}\in[-\frac{\pi}{2},\frac{\pi}{2}]$ and meridional $\varphi^{\dag}\in[0,2\pi)$  angles (given by the analogous of \eqref{rot coord}), the coordinates of a point $P$ on the Earth are
\begin{equation}\label{rdag}
	\Vec{OP}:=\mathbf{r}= r\cos\td\cos\pd \ea^{\dag}+r\cos\td\sin\pd \eb^{\dag}+r\sin\td \ec^{\dag}.
\end{equation}
Therefore, using \eqref{4.3} and equaling \eqref{r} and \eqref{rdag}, it follows that
\begin{equation}\label{rot identities}
	\left\{\begin{aligned}
		&\cos\theta^{\dag}\cos\varphi^{\dag}=\cos\theta\sin\varphi,\\
		&\cos\theta^{\dag}\sin\varphi^{\dag}=\sin\theta,\\
		&\sin\theta^{\dag}=\cos\theta\cos\varphi.
	\end{aligned}\right.
\end{equation}
In this new coordinate system the North Pole has coordinates $\pd=\frac{\pi}{2},\ \td=0$, and the solution of the system \eqref{rot identities} for the Arctic Ocean is 
\begin{equation}
	\left\{\begin{aligned}
		&\pd=\cot^{-1}(\sin\varphi \cot\theta) \in (0,\pi),\\
		&\td=\sin^{-1}(\cos\varphi \cos\theta) \in \left(-\frac{\pi}{2},\frac{\pi}{2}\right).
	\end{aligned}\right.
\end{equation}
The unit basis vectors in the rotated spherical coordinates $(\pd, \td, r)$ are  $(\ep^{\dag}, \et^{\dag}, \er)$, given by
\begin{equation}
	\left\{ \begin{array}{ll}
		\ep^{\dag}=-\sin\varphi^{\dag} \ea^{\dag}+\cos\varphi^{\dag} \eb^{\dag},\\
		\et^{\dag}=-\cos\varphi^{\dag}\sin\theta^{\dag} \ea^{\dag}-\sin\varphi^{\dag}\sin\theta^{\dag} \eb^{\dag}+\cos\theta^{\dag} \ec^{\dag},\\
		\er^{\dag}=\cos\varphi^{\dag}\cos\theta^{\dag} \ea^{\dag}+\sin\varphi^{\dag}\cos\theta^{\dag} \eb^{\dag}+\sin\theta^{\dag} \ec^{\dag},
	\end{array}\right.
	\label{base sferiche dag}
\end{equation}
and the corresponding velocity components are $\bu=(u^{\dag}, v^{\dag}, w^{\dag})$. The two coordinate systems are depicted in Figure \ref{coordinates}.\\
\subsection{Equations of motion}
Away from the boundary layers, the friction/viscous effects are negligible, so the fluid can be considered ideal \cite{CJBOOK}. The governing equations for a rotating incompressible ideal fluid are, in vectorial coordinate-free form, written as 
\begin{equation}\label{V G}
	\left\{ \begin{aligned}
		& \frac{D\bu}{Dt}+2\mathbf{\Omega}\times\bu+\mathbf{\Omega}\times(\mathbf{\Omega}\times \mathbf{r})=\frac{\mathbf{G}-\nabla p}{ \rho },\\
		& \frac{D\rho}{Dt}=0,\\
		&  \nabla\cdot \bu=0,
	\end{aligned}\right.
\end{equation}
where $\frac{D}{Dt}=\left(\frac{\partial}{\partial t} + \bu\cdot\nabla\right)$ is the material derivative, $\bu$ is the velocity vector, $\rho$ is the fluid density, $\mathbf{G}$ is the body force per unit volume, $p$ is the pressure, $\mathbf{\Omega}$ the rotation vector and $\mathbf{r}$ the position vector \cite{ConstantinBook},  \cite{Morita}. In the subsequent analysis, we will consider three constant densities $\rho_0$, $\rho_1$ and $\rho_2$, with $\rho_0<\rho_1<\rho_2$, in different regions of the flow; in each layer the fluid can be considered incompressible.\\
The three equations in \eqref{V G} correspond to, respectively, the momentum equation, the mass conservation equation, and the incompressibility condition.\\
In the new rotated spherical coordinates the Euler equations are written as (see \cite{MIO} for a complete derivation of the Navier-Stokes equations in the rotated spherical coordinate system, from which the incompressible Euler equations reduces to, when considering constant density and zero viscosity)
\begin{equation}\label{Euler rotated}
	\begin{aligned}
		& \left[ \frac{\partial} {\partial t}+ \frac{u^{\dag}}{r \cos \theta^{\dag}} \frac{\partial }{\partial \varphi^{\dag}} + \frac{v^{\dag}}{r} \frac{\partial }{\partial \theta^{\dag}} + w^{\dag}\frac{\partial }{\partial r} \right]\begin{pmatrix}
			u^{\dag} \\
			v ^{\dag}\\
			w^{\dag}\\
		\end{pmatrix}+\\
        &+\frac{1}{r}\begin{pmatrix}-u^{\dag}v^{\dag}\tan \theta^{\dag} + u^{\dag}w ^{\dag}\\
			u^{\dag 2}\tan\theta^{\dag}+v^{\dag}w^{\dag}\\
			-u^{\dag 2}-v^{\dag 2}
		\end{pmatrix}+2\Omega\begin{pmatrix}
			-v^{\dag}\sin\varphi^{\dag}\cos\theta^{\dag}-w^{\dag}\sin\varphi^{\dag}\sin\theta^{\dag}\\
			u^{\dag}\sin\varphi^{\dag}\cos\theta^{\dag}-w^{\dag}\cos\varphi^{\dag}\\
			u^{\dag} \sin\varphi^{\dag}\sin\theta^{\dag} +v^{\dag} \cos\varphi^{\dag}
		\end{pmatrix}+\\
        &+r \Omega^2\begin{pmatrix}
			\sin\varphi^{\dag}\cos\varphi^{\dag}\cos\theta^{\dag}\\
			-\sin^2\varphi^{\dag} \sin\theta^{\dag}\cos\theta^{\dag}\\
			-\cos^2\varphi^{\dag}\cos^2\theta^{\dag} -\sin^2\theta^{\dag}
		\end{pmatrix} =-\frac{1}{\rho}\begin{pmatrix}
			\frac{1}{r \cos \theta^{\dag}} \frac{\partial p}{\partial \varphi^{\dag}} \\
			\frac{1}{r} \frac{\partial p}{\partial \theta^{\dag}} \\
			\frac{\partial p}{\partial r}\\
		\end{pmatrix} -\begin{pmatrix}
			0\\
			0\\
			g\\
		\end{pmatrix},
	\end{aligned}
\end{equation}
and the mass conservation condition 
\begin{equation}\label{mass rotated}
	\frac{\partial \rho} {\partial t}+ \frac{u^{\dag}}{r \cos \theta^{\dag}} \frac{\partial \rho}{\partial \varphi^{\dag}} + \frac{v^{\dag}}{r} \frac{\partial \rho}{\partial \theta^{\dag}} + w^{\dag}\frac{\partial \rho}{\partial r} =0,
\end{equation}
with $\rho=\rho_i$, $i=0,1,2$, depending on region of the flow, and the incompressibility condition is given by
\begin{equation}\label{inc rot}
	\frac{1}{r \cos \theta^{\dag}} \frac{\partial u^{\dag}}{\partial \varphi^{\dag}} + \frac{1}{r } \frac{\partial v^{\dag} }{\partial \theta^{\dag}}  -\frac{v^{\dag}}{r}\tan \theta^{\dag}+  \frac{\partial w^{\dag}}{\partial r} +\frac{2}{r}w^{\dag}=0.
\end{equation}
Moreover, as $\mathbf{\Omega}\times(\mathbf{\Omega}\times \mathbf{r})=\frac{1}{2}\nabla(\Omega^2\ell^2)$
, where $\ell$ is the distance of the point from the axis of rotation $\ec$, 
we can redefine the pressure as
\begin{equation}\label{pressure}
	P=p-\frac{1}{2}\rho\ r^2\Omega^2 (\cos^2\varphi^{\dag}\cos^2\theta^{\dag} + \sin^2\theta^{\dag}).
\end{equation}
With this redefined pressure, the Euler equations \eqref{Euler rotated} read as
\begin{equation}\label{Euler rotated 2}
	\begin{aligned}
		& \left[ \frac{\partial} {\partial t}+ \frac{u^{\dag}}{r \cos \theta^{\dag}} \frac{\partial }{\partial \varphi^{\dag}} + \frac{v^{\dag}}{r} \frac{\partial }{\partial \theta^{\dag}} + w^{\dag}\frac{\partial }{\partial r} \right]\begin{pmatrix}
			u^{\dag} \\
			v ^{\dag}\\
			w^{\dag}\\
		\end{pmatrix}+\frac{1}{r}\begin{pmatrix}-u^{\dag}v^{\dag}\tan \theta^{\dag} + u^{\dag}w ^{\dag}\\
			u^{\dag 2}\tan\theta^{\dag}+v^{\dag}w^{\dag}\\
			-u^{\dag 2}-v^{\dag 2}
		\end{pmatrix}\\
		&+2\Omega\begin{pmatrix}
			-v^{\dag}\sin\varphi^{\dag}\cos\theta^{\dag}-w^{\dag}\sin\varphi^{\dag}\sin\theta^{\dag}\\
			u^{\dag}\sin\varphi^{\dag}\cos\theta^{\dag}-w^{\dag}\cos\varphi^{\dag}\\
			u^{\dag} \sin\varphi^{\dag}\sin\theta^{\dag} +v^{\dag} \cos\varphi^{\dag}
		\end{pmatrix}=-\frac{1}{\rho}\begin{pmatrix}
			\frac{1}{r \cos \theta^{\dag}} \frac{\partial P}{\partial \varphi^{\dag}} \\
			\frac{1}{r} \frac{\partial P}{\partial \theta^{\dag}} \\
			\frac{\partial P}{\partial r}\\
		\end{pmatrix} -\begin{pmatrix}
			0\\
			0\\
			g\\
		\end{pmatrix}.
	\end{aligned}
\end{equation}
\subsubsection{Standard approximations in geophysical fluid dynamics}
In this subsection, we review the most common approximations in geophysical fluid dynamics and adapt them to the rotated spherical coordinate system we are using for the Arctic Ocean.
\paragraph{\bf Tangent plane approximation} We approximate the region of the Earth around a point $P$ (with rotated spherical coordinates $P=(\varphi^{\dag}_0, \theta^{\dag}_0, R)$, where $R$ is the radius of the Earth) by the flat geometry of the tangent plane. The relation between the coordinates $(X^{\dag}, Y^{\dag}, Z^{\dag})$ referred to the basis $(\mathbf{e}_1^{\dag}, \mathbf{e}_2^{\dag}, \mathbf{e}_3^{\dag}) $ and rotated spherical coordinates $(\varphi^{\dag}, \theta^{\dag}, r)$ referred to the basis $(\ep^{\dag}, \et^{\dag}, \er^{\dag}) $ is given by
\begin{equation}
	\left\{ \begin{array}{ll}
		X^{\dag}=r \cos\theta^{\dag} \cos \varphi^{\dag},\\
		Y^{\dag}=r \cos\theta^{\dag} \sin \varphi^{\dag},\\
		Z^{\dag}=r \sin\theta ^{\dag}.
	\end{array}\right.
	\label{tg1}
\end{equation}
It is well know that the region around $P$ can be approximated by a tangent plane. The sphere of radius $R$ is given by $\mathbf{S}^2_R=\{(X^{\dag}, Y^{\dag}, Z^{\dag})\in\mathbb{R}^3\,s.t.\, F(X^{\dag}, Y^{\dag}, Z^{\dag}):=X^{\dag^2}+Y^{\dag^2}+ Z^{\dag^2}-R^2=0\}$, hence the normal vector to the sphere at $P$ is given by (using \eqref{tg1})
\begin{equation}
\begin{aligned}
    \boldsymbol{\mathfrak{n}}&=\left(\nabla F\right)(P)=2(X_P^{\dag}, Y_P^{\dag}, Z_P^{\dag})=\\
    &=2R(\cos\theta_0\cos\varphi_0,\, \cos\theta_0\sin\varphi_0,\, \sin\theta_0).
\end{aligned}	
\end{equation}
The vector $\boldsymbol{\mathfrak{n}}$ is (up to normalizing) equal to the base vector $\er^{\dag}(P)$ (see equation \eqref{base sferiche dag}), and, since the tangent plane to the sphere at the point $P$ is the geometric locus of points $M$ such that $\Vec{PM}\cdot\boldsymbol{\mathfrak{n}}=\Vec{PM}\cdot\er(P)=0$, it's an immediate consequence that a basis for the tangent plane at $P$ is $(\ep^{\dag}(P), \et^{\dag}(P))$.\\
On the tangent plane we define the following local coordinates
\begin{equation}\label{tg change}
	\left\{\begin{aligned}
		&\x=R \cos\theta^{\dag}_0 (\varphi^{\dag}-\varphi^{\dag}_0),\\
		& \y=R (\theta^{\dag}-\theta^{\dag}_0),\\
	\end{aligned}\right.
\end{equation}
therefore, computing
\begin{equation}\label{nabla tg plane}
	\nabla \x = R\cos\theta^{\dag}_0\ \ep^{\dag}(P),\\ 
	\qquad\nabla \y= R\ \et^{\dag}(P),
\end{equation}
we obtain the following basis 
\begin{equation}
	\mathbf{e}_{\x}=\frac{\nabla \x }{|\nabla \x|}=\ep(P),\qquad
	\mathbf{e}_{\y}=\frac{\nabla \y }{|\nabla \y|}=\et(P),
\end{equation}
proving that $(\ep^{\dag}(P), \et^{\dag}(P) )$ is in fact a basis for the tangent plane at $P$.\\
The concept of tangent plane is always used implicitly for the horizontal variables when adopting an $f$-plane approximation.\\ 
The partial derivatives with respect to $(\x, \y) $  are given by
\begin{equation}\label{derivate prime tg plane}
	\frac{\partial}{\partial \x}=\frac{1}{R\cos\theta_0}\frac{\partial}{\partial \varphi^{\dag}},\qquad     \frac{\partial}{\partial \y}=\frac{1}{R}\frac{\partial}{\partial \theta^{\dag}}.
\end{equation}
\paragraph{\bf Traditional approximation}
In this paragraph, we review the ideas leading to the so-called traditional approximation, adapting them to the rotated spherical coordinates. Namely, we will construct this approximation starting from \eqref{Euler rotated 2}. Note that, even if the velocities $u^{\dag}, v^{\dag}, w^{\dag}$ are defined differently from the ones relative to the classical spherical coordinate system, $u^{\dag}, v^{\dag}$ are still the horizontal and $w^{\dag}$ the vertical components of the velocity vector $\bu$, hence the observations about their magnitude, on which the traditional approximation is based, still apply (see \cite{Vallis}). For a review of the traditional approximation of the governing equations in classical spherical coordinates, see also \cite{MarshallPlumb}.\\
The first step involves neglecting metric terms \\
\begin{equation}\frac{1}{r}\left(-u^{\dag}v^{\dag}\tan \theta^{\dag} + u^{\dag}w ^{\dag},\,
u^{\dag 2}\tan\theta^{\dag}+v^{\dag}w^{\dag},\,
-u^{\dag 2}-v^{\dag 2}
\right),\end{equation} which represent the effect of curvature in spherical coordinates, in \eqref{Euler rotated 2}, and the terms in \eqref{mass rotated} and \eqref{inc rot} which involve a velocity multiplied by $\frac{1}{r}$.\\ 
Because of the thinnes of the atmosphere and ocean, vertical velocities (typically $\leq 0.01\, \mathrm{ms^{-1}}$) are much less than horizontal velocities by a factor of $10^{-4}$ on average, therefore we may neglect the term involving $w^{\dag}$ in the horizontal components (namely the first two) of the Coriolis acceleration
\begin{equation}   
2\mathbf{\Omega} \times \mathbf{u}=2\Omega\begin{pmatrix}   
-v^{\dag}\sin\varphi^{\dag}\cos\theta^{\dag}-w^{\dag}\sin\varphi^{\dag}\sin\theta^{\dag}\\
u^{\dag}\sin\varphi^{\dag}\cos\theta^{\dag}-w^{\dag}\cos\varphi^{\dag}\\
u^{\dag} \sin\varphi^{\dag}\sin\theta^{\dag} +v^{\dag} \cos\varphi^{\dag}
\end{pmatrix}.
\end{equation} Moreover, in the atmosphere the typical scale of the horizontal velocities is  $|u^{\dag}|,\, |v^{\dag}|\approx 10\, \mathrm{ms^{-1}}$, and even less in the ocean, giving that $|2\Omega (u^{\dag}+v^{\dag})|\approx 3\cdot 10^{-3}\, \mathrm{ms^{-2}}$, is negligible compared to gravity acceleration $g\approx 9.8\,\mathrm{ms^{-2}}$. Therefore, the Coriolis acceleration can be approximated as
\begin{equation}
    2\mathbf{\Omega} \times \mathbf{u}\approx2\Omega\begin{pmatrix}
        -v^{\dag}\sin\varphi^{\dag}\cos\theta^{\dag}\\
u^{\dag}\sin\varphi^{\dag}\cos\theta^{\dag}\\
0
    \end{pmatrix}.
\end{equation}
These two approximations related to the Coriolis term $2\mathbf{\Omega} \times \mathbf{u}$, the first involving the two horizontal components of the equation, and the second involving the vertical component where the gravity acceleration is present,  cannot be done independently of each other. Instead, they must be made together, otherwise the resulting equations fail to conserve energy \cite{Salmon}.\\
Lastly, as the ocean is shallow compared to Earth's radius, we can write $r=R+\z$, with $R$ being the Earth's radius and $\z$ increasing in the radial direction. This idea is fundamental for the so-called `thin-shell' approximation (see e.g \cite{CJ2023} or \cite{JBook}). The coordinate $r$ is then replaced by $R$, except in the derivatives. For example $\frac{1}{r^2}\frac{\partial r^2 w^{\dag}}{\partial r}$ becomes $\frac{\partial w^{\dag}}{\partial \z}$. With the above arguments, the Euler, mass conservation and incompressibility equations reduce to
\begin{equation}\label{Euler rotated approximated}
\left\{\begin{aligned}
	& \left[ \frac{\partial} {\partial t}+ \frac{u^{\dag}}{R \cos \theta^{\dag}} \frac{\partial }{\partial \varphi^{\dag}} + \frac{v^{\dag}}{R} \frac{\partial }{\partial \theta^{\dag}} + w^{\dag}\frac{\partial }{\partial \z} \right]\begin{pmatrix}
		u^{\dag} \\
		v ^{\dag}\\
		w^{\dag}\\
	\end{pmatrix}+\\
    &\qquad +2\Omega\begin{pmatrix}
		-v^{\dag}\sin\varphi^{\dag}\cos\theta^{\dag}\\
		u^{\dag}\sin\varphi^{\dag}\cos\theta^{\dag}\\
		0
	\end{pmatrix}=-\frac{1}{\rho}\begin{pmatrix}
		\frac{1}{R \cos \theta^{\dag}} \frac{\partial P}{\partial \varphi^{\dag}} \\
		\frac{1}{R} \frac{\partial P}{\partial \theta^{\dag}} \\
		\frac{\partial P}{\partial \z}\\
	\end{pmatrix} -\begin{pmatrix}
		0\\
		0\\
		g\\
	\end{pmatrix},\\
	&\,\frac{\partial \rho} {\partial t}+ \frac{u^{\dag}}{R \cos \theta^{\dag}} \frac{\partial \rho}{\partial \varphi^{\dag}} + \frac{v^{\dag}}{R} \frac{\partial \rho}{\partial \theta^{\dag}} + w^{\dag}\frac{\partial \rho}{\partial \z} =0,\\
	&\,\frac{1}{R \cos \theta^{\dag}} \frac{\partial u^{\dag}}{\partial \varphi^{\dag}} + \frac{1}{R } \frac{\partial v^{\dag} }{\partial \theta^{\dag}}  +  \frac{\partial w^{\dag}}{\partial \z} =0,
\end{aligned}\right.
\end{equation}
respectively.
\paragraph{\bf $f$-plane approximation in the rotated spherical coordinates framework}
Usually, the Coriolis parameter is defined by \( f =2\Omega\sin\theta\), where $\theta \in \left[-\frac{\pi}{2}, \frac{\pi}{2}\right]$ is the classical angle of latitude (see \cite{Morita}), but, due to the relation $\sin\theta=\cos\theta^{\dag}\sin\varphi^{\dag}$ in \eqref{rot identities} we can write 
\begin{equation}f=2\Omega\sin\theta=2\Omega\cos\theta^{\dag}\sin\varphi^{\dag},\end{equation}
and it follows that
\[\frac{\partial f}{\partial \varphi^{\dag}}=2\Omega\cos\theta^{\dag}\cos\varphi^{\dag},\qquad \frac{\partial f}{\partial \varphi^{\dag}}=-2\Omega\sin\theta^{\dag}\sin\varphi^{\dag}.\] 
Using a Taylor expansion at the first order, we can write
\begin{equation}
\begin{aligned}
	f&=f_0+\frac{\partial f (\varphi_0^{\dag},\theta^{\dag}_0)}{\partial \varphi^{\dag}}\left(\varphi^{\dag}-\varphi^{\dag}_0\right)+\frac{\partial f (\varphi_0^{\dag},\theta^{\dag}_0)}{\partial \varphi^{\dag}}\left(\theta^{\dag}-\theta^{\dag}_0\right)+\mathcal{O}\left(\varphi^{\dag^2},\theta^{\dag^2}\right)=\\    &=f_0+\mathcal{O}\left(\varphi^{\dag},\theta^{\dag}\right),
\end{aligned}
\end{equation}
where $f_0=2\Omega\cos\theta_0^{\dag}\sin\varphi_0^{\dag}$.\\
As described previously, in the rotated spherical coordinates framework,   a small region around a specific point of coordinates \( \theta^{\dag}_0, \varphi^{\dag}_0, R\), can be described by taking a tangent plane approximation (for the horizontal coordinates),  given by \eqref{tg change}
\begin{equation}
\left\{\begin{aligned}
	&\x=R \cos\theta^{\dag}_0 (\varphi^{\dag}-\varphi^{\dag}_0),\\
	& \y=R (\theta^{\dag}-\theta^{\dag}_0),\\
\end{aligned}\right.
\end{equation} 
so that the \( \x \)-axis points toward the North Pole, the \( \y \)-axis points toward the Null Island, while the vertical \( \z \)-axis is assume to point upward. As for the classical $f$-plane approximation, the Coriolis parameter $f$ is assumed to remain constant and equal to $f_0$ within the localized region. As an $f$-plane approximation is implicitly based on a tangent plane approximation, the error committed by adopting this approximation is of order of $\mathcal{O}\left(\frac{\x}{R},\frac{\y}{R}\right).$\\
The Euler, mass conservation and incompressibility equations in the $f$-plane approximation are given by
\begin{equation}\label{Euler f-plane}
\left\{\begin{aligned}
	& \left[ \frac{\partial} {\partial t}+  u^{\dag} \frac{\partial }{\partial \x} +  v^{\dag}\frac{\partial }{\partial \y} + w^{\dag}\frac{\partial }{\partial \z} \right]\begin{pmatrix}
		u^{\dag} \\
		v ^{\dag}\\
		w^{\dag}\\
	\end{pmatrix}+\begin{pmatrix}
		-fv^{\dag}\\
		fu^{\dag}\\
		0
	\end{pmatrix}=-\frac{1}{\rho}\begin{pmatrix}
		\frac{\partial P}{\partial \x} \\
		\frac{\partial P}{\partial \y} \\
		\frac{\partial P}{\partial \z}\\
	\end{pmatrix} -\begin{pmatrix}
		0\\
		0\\
		g\\
	\end{pmatrix},\\
	&\frac{\partial \rho} {\partial t}+  u^{\dag} \frac{\partial \rho}{\partial \x} +  v^{\dag}  \frac{\partial \rho}{\partial \y} + w^{\dag}\frac{\partial \rho}{\partial \z} =0,\\
	&\frac{\partial u^{\dag}}{\partial \x} + \frac{\partial v^{\dag} }{\partial \y}  +  \frac{\partial w^{\dag}}{\partial \z} =0.
\end{aligned}\right.
\end{equation}
These equations hold for general $\rho=\rho(x,y,z)$, but, as anticipated, we will consider three constant densities $\rho_0,\ \rho_1,\ \rho_2$; consequently the mass conservation equation (namely the second of \eqref{Euler f-plane}) will always satisfied. Across the interfaces, where the density changes, dynamic (pressure continuity) and kinematic (no particle mixing) boundary conditions will imposed, as will be detailed in Section \ref{sec3}.

\subsection{North Pole and coordinates relative to the TDC}
\begin{figure}[ht]
\centering
\includegraphics[width=.6\textwidth]{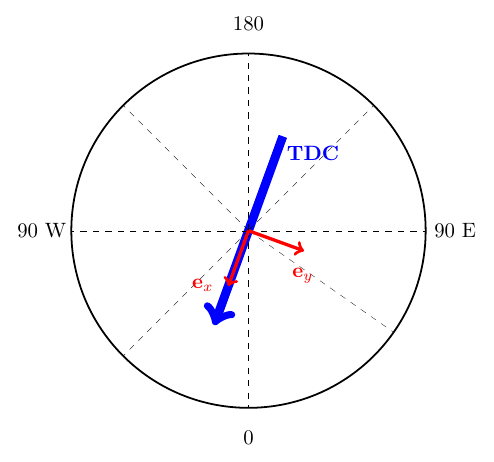}
\caption{Depiction of the basis at the North Pole. In reality, the angle $\gamma$ is bigger than the one depicted here (close to $90^{\circ}$). However, for graphical clarity reasons, we did it smaller ($70^{\circ}).$}
\label{Basis NorthPole}
\end{figure}

The use of the newly defined, rotated spherical coordinates \((\pd, \td, r)\), along with the corresponding governing equations, is particularly applicable when analyzing flows in regions centred at the poles, where standard spherical coordinates are not defined.\\
Our analysis focus on a region centred at the  North Pole, having coordinates $\pd_0=\frac{\pi}{2},\ \td_0=0$ (and $r=R$). This imply that $f_0=2\Omega$ and the tangent plane coordinates are given by
\begin{equation}
\left\{\begin{aligned}
	&\x=R \left(\varphi^{\dag}-\frac{\pi}{2}\right),\\
	& \y=R \theta^{\dag},\\
\end{aligned}\right.
\end{equation} 
with basis given by $\ep^{\dag}(\mathfrak{N}), \et^{\dag}((\mathfrak{N}))$, with $\mathfrak{N}$ representing the North Pole. See Figure \ref{Basis NorthPole} for a depiction of the basis at the North Pole.\\
\\
The final step involves defining a new basis $\mathbf{e}_x, \mathbf{e}_y$ for the tangent plane, by simply rotating anticlockwise the basis $\ep^{\dag}(\mathfrak{N}), \et^{\dag}((\mathfrak{N}))$ of a certain angle $\gamma$, in order to align the first basis vector with the Transpolar Drift Current (this angle would be around $80^{\circ}$ or even $90^{\circ}$).\\
These unit vectors are given by (see Figure \ref{Basis NorthPole})
\begin{equation}\label{tdc basis}
\left\{\begin{aligned}
    &\mathbf{e}_x= \cos\gamma \ep^{\dag}(\mathfrak{N})+\sin\gamma \et^{\dag}((\mathfrak{N})),\\
&\mathbf{e}_y=-\sin\gamma\ep^{\dag}(\mathfrak{N})+\cos\gamma\et^{\dag}((\mathfrak{N})).
\end{aligned}\right.
\end{equation}
Finally, observe that, as $\mathbf{\Omega}\times(\mathbf{\Omega}\times \mathbf{r})=\frac{1}{2}\nabla(\Omega^2\ell^2)$ where $\ell$ is the distance from the axis of rotation $\ec$, this will not change when adopting this new basis, hence there is no need to modify the expression for the pressure \eqref{pressure}.\\
The governing equations with respect to the new basis are therefore given by
\begin{equation}\label{Euler f plane artic}
\left\{\begin{aligned}
	&\frac{\partial u} {\partial t} + u \frac{\partial u}{\partial x} + v \frac{\partial u }{\partial y} + w\frac{\partial u}{\partial z} 
	-fv=
	-\frac{1}{\rho}
	\frac{\partial P}{\partial x}, \\
	&\frac{\partial v} {\partial t} + u \frac{\partial v}{\partial x} + v \frac{\partial v }{\partial y} + w\frac{\partial v}{\partial z} +
	f u=
	-\frac{1}{\rho}
	\frac{\partial P}{\partial y}, \\
	&\frac{\partial w} {\partial t} + u \frac{\partial w}{\partial x} + v \frac{\partial w }{\partial y} + w\frac{\partial w}{\partial z} +g
	=
	-\frac{1}{\rho}
	\frac{\partial P}{\partial z} 	,
\end{aligned}\right.
\end{equation}
\begin{equation}\label{cont eq}
\frac{\partial u}{\partial x}+\frac{\partial v}{\partial y}+\frac{\partial w}{\partial z}=0,
\end{equation}
where $(u,v,w)$ are the velocity component associated to the basis $\left(\mathbf{e}_x, \mathbf{e}_y, \er(\mathfrak{N})\right)$, and with $\rho$ equal $\rho_0,\ \rho_1$ or $\rho_2$ depending on the layer of the fluid we are considering.\\
For consistency of notation, we have used $z$ in place of $\z$.
\section{Description of the flows}\label{sec3}
The simplified model under consideration consists of three layers of constant density, denoted by $\rho_0$, $\rho_1$, and $\rho_2$, with $\rho_0<\rho_1<\rho_2$. The surface mixed layer (SML), composed of cold and relatively fresh water above the halocline, has density $\rho_0$. The halocline itself is characterised by density $\rho_1$, while the denser, warmer, and saltier Atlantic Water (AW) beneath the halocline has density $\rho_2$.\\
As depth increases, turbulent effects generated by momentum exchange at the water surface rapidly decay. This attenuation is even more pronounced in the presence of sea ice, a defining feature of the oceanic region considered here. For a detailed analysis of wind--ice--water interactions at the Arctic Ocean surface, we refer to \cite{MIO} where a solution for  nonlinear wind-induced ice-drift currents is found, featuring the superposition of near-inertial oscillations, a mean Ekman flow and a geostrophic current\footnote{We remark that, for studies of ice--sea or air--sea interactions, the appropriate governing equations are the Navier--Stokes equations rather than the Euler equations, which are used in this work, as we focus on the deeper parts of the ocean.}.\\
We therefore assume that just above the lower boundary of the SML, denoted by $\eta_1$, there exists a (possibly thin) layer in which the fluid flows in the direction of the Transpolar Drift Current (TDC) with a mean velocity of approximately $0.1\,\mathrm{m\,s^{-1}}$. We denote the upper boundary of this layer by $\eta_0$.\\
At the interface $\eta_0$, we impose both the dynamic boundary condition (continuity of pressure) and the kinematic boundary condition (no mass exchange across the interface). The latter reflects the assumption that turbulence---and hence mixing---is absent in this region. As will become apparent from the analysis, the precise expression of $\eta_0$ does not affect our analysis.\\
We further assume that the fluid below the halocline is practically motionless. The lower boundary of the halocline is denoted by $\eta_2$.\\
In addition to the governing equations \eqref{Euler f plane artic} and \eqref{cont eq}, the solution is required to satisfy the dynamic boundary condition, ensuring pressure continuity across the interfaces $\eta_1$ and $\eta_2$, as well as the kinematic boundary condition
\begin{equation}\label{KBC}
w=\frac{\partial\eta_i}{\partial t}
 +u\frac{\partial\eta_i}{\partial x}
 +v\frac{\partial\eta_i}{\partial y}
\quad \text{on } z=\eta_i,\quad i=1,2,
\end{equation}
which prevents the exchange of fluid particles between adjacent layers (see \cite{ConstantinBook}). This condition reflects the strong stratification of the halocline, which inhibits direct contact between the warmer, saltier AW and the sea ice above.\\
Owing to the distinct physical properties of each layer, we analyse the flow in each region separately, proceeding from the deepest layer upward.

\subsection{The deep motionless layer beneath the halocline}

In the region $z<\eta_2(x-\csf t,y)$  the water is in hydrostatic state $u=v=w=0$, hence \eqref{Euler f plane artic} reads as
\begin{equation}
\left\{	\begin{aligned}
	&
	\frac{\partial P}{\partial x}=0, \\
	&	
	\frac{\partial P}{\partial y}= 0,\\
	&	\frac{\partial P}{\partial z} 	=-\rho_2 g,
\end{aligned}\right.
\end{equation}
giving, for the pressure, 
\begin{equation}\label{pressure under halocline}
P(x,y,z)=P_2-\rho_2 g z ,
\end{equation}
with $P_2\in\mathbb{R}$. Being hydrostatic, the flow is irrotational 
and satisfies the continuity equation \eqref{cont eq}.

\subsection{The halocline layer  $\H(t)$ }
For $\eta_2(x-\csf t,y)\leq z \leq \eta_1(x-\csf t,y)$ we seek an explicit solution of \eqref{Euler f plane artic}, \eqref{cont eq} and \eqref{KBC}, fulfilling also the dynamic boundary condition, in the Lagrangian formalism (see \cite{Abrashkin} for a review of the Lagrangian description of fluid flows): at time $t$ we specify the positions of the fluid particles in terms of the labeling variables $(q,\ r,\ s)$ by
\begin{equation}\label{Lag}
\left\{	\begin{aligned}
	&x=q-be^{-ms}\sin(k(q-\csf t)),\\
	&y=r-de^{-ms}\cos(k(q-\csf t)),\\
	&z=-d_0+s-ae^{-ms}\cos(k(q-\csf t)).
\end{aligned}\right.
\end{equation}
The constant $k=\frac{2\pi}{L}>0$ is the wave number corresponding to the wavelength $L$, and we assume $m>0$  and $a>0$, while the labeling variables are chosen so that
\begin{equation}
(q,r,s) \in [-\mathrm{q}_0,\mathrm{q}_0]\times[-\mathrm{r}_0,\mathrm{r}_0]\times [\mathrm{s}_-(r),\mathrm{s}_+(r) ],
\end{equation}
where $s=\mathrm{s}_-(r)\geq \mathrm{s}^*>0$ represents $\eta_2$, $s=\mathrm{s}_+(r)>\mathrm{s}_-(r)$ represents $\eta_1$, while $d_0-\mathrm{s_-}$ is the mean depth of the halocline base. Moreover, $\mathrm{q}_0$ and $\mathrm{r}_0$ are chosen so that the area under consideration is contained in the Arctic Ocean basin.\\
It will be shown that \eqref{Lag}, representing waves with crests parallel to the $y$-axis and propagating in the direction of the $x$-axis, is a solution of the Euler and incompressibility equations.\\
The particle motion in \eqref{Lag} describes trochoidal orbits, namely the path of a fixed point on a circle of radius $be^{-ms}$ and centred at $(q,r,s-d_0)$,  in a plane that is at an angle $\arctan(-d/a)$ with respect to the vertical axis.\\
Anticipating the relation $a^2+d^2=b^2$, cf. \eqref{adb}, and recalling that a circle in three dimension in uniquely defined by six numbers (three for the circle's center, one for the radius and two for the orientation of the unit vector normal to the plane of the circle), namely:
\begin{equation}\label{cerchio}
    \mathrm{circle}=\left\{\mathbf{C}+\mathcal{R} \cos{\theta}\boldsymbol{\mu}+\mathcal{R} \cos{\theta}\boldsymbol{\nu}\times\boldsymbol{\mu}\right\}_{\theta \in[0,2\pi)},
\end{equation}
where $\mathbf{C}$ is the center of the circle, $\mathcal{R}$ its radius, $\boldsymbol{\nu}$ is the unit vector normal to the plane of the circle, $\boldsymbol{\mu}$ is a vector orthogonal to $\boldsymbol{\nu}$ and $\theta$ is the angular position on the circle, it is immediate to see that we recover \eqref{Lag} by setting
\begin{equation}\label{cerchioParameters}
    \begin{aligned}
       & \mathbf{C}=(q,r,s-d_0),   \qquad\mathcal{R}=be^{-ms},\qquad    \theta=\tau=k(q-\csf t),\\
        &\boldsymbol{\nu}=\left(0, \frac{a}{b}, \frac{-d}{b}\right),\qquad
        \boldsymbol{\mu}=\left(0, \frac{-d}{b}, \frac{a}{b}\right).      
    \end{aligned}
\end{equation}
The Jacobian of the map \eqref{Lag} relating the particle positions with the Lagrangian labeling variables is given by
\begin{equation}\label{18}
\begin{aligned}
&\left(\frac{\partial(x,y,z)}{\partial(q,r,s)}\right)=\begin{pmatrix}
	\frac{\partial x}{\partial q} & \frac{\partial y}{\partial q} & \frac{\partial z}{\partial q} \\
	\frac{\partial x}{\partial r} & \frac{\partial y}{\partial r} & \frac{\partial z}{\partial r}	\\	
	\frac{\partial x}{\partial s} & \frac{\partial y}{\partial s} & \frac{\partial z}{\partial s}
\end{pmatrix}=\\
&\qquad \quad=\begin{pmatrix}
	1-kbe^{-ms}\cos\tau & kde^{-ms}\sin\tau & kae^{-ms}\sin\tau \\
	0 & 1 & 	0\\
	mbe^{-ms}\sin\tau & mde^{-ms}\cos\tau &1+mae^{-ms}\cos\tau
\end{pmatrix},
\end{aligned}
\end{equation}
with
\begin{equation}
\tau=k(q-\csf t),
\end{equation}
and its determinant is expressed by
\begin{equation}\label{jacobian}
J=1+(am-bk)e^{-ms}\cos\tau-kmab\,e^{-2ms}.
\end{equation}
The flow is incompressible, namely \eqref{cont eq} holds, if $J$ is time-independent and non-zero (see \cite{ConstantinBook}), which requires
\begin{equation}\label{9Mc}
am=bk,
\end{equation}
thus providing
\begin{equation}\label{jacobian2}
J=1-m^2a^2\,e^{-2ms}\not=0.
\end{equation}
The above condition that $J\not=0$ implies that \eqref{Lag} is a local diffeomorphic change of coordinates by the inverse function theorem. Since $s\geq\mathrm{s}^*>0$,
\begin{equation}\label{<1}
a^2m^2\,e^{-2m\mathrm{s}^*}<1,
\end{equation} 
and, as $ae^{-m\mathrm{s}^*}$ is the amplitude of a wave at some $r$, it is possible to find an upper bound for the vertical amplitude:
\begin{equation}\label{amax}
a_{\mathrm{max}}=\frac{1}{m}.
\end{equation}
Morevoer, due to \eqref{<1}, it is evident that 
\begin{equation}
J=1-m^2a^2\,e^{-2ms}>0.
\end{equation}
The velocity and acceleration of a particle can be computed using \eqref{Lag}, giving respectively
\begin{equation}\label{20 e 21}
\left\{\begin{aligned}
	&u=\frac{	Dx}{Dt}=k\csf b\, e^{-ms}\cos\tau,\\
	&v=\frac{	Dy}{Dt}=-k\csf d\, e^{-ms}\sin\tau,\\
	&w=\frac{	Dz}{Dt}=-k\csf a\, e^{-ms}\sin\tau,
\end{aligned}\right.
\quad\text{and} \quad
\left\{\begin{aligned}
	&\frac{	Du}{Dt}=k^2\csf^2 b\, e^{-ms}\sin\tau,\\
	&\frac{	Dv}{Dt}=k^2\csf^2 d\, e^{-ms}\cos\tau,\\
	&\frac{	Dw}{Dt}=k^2\csf^2 a\, e^{-ms}\cos\tau,
\end{aligned}\right.
\end{equation}
therefore, the Euler equations \eqref{Euler f plane artic} rewritten in a more compact form as
\begin{equation}\label{4}
\left\{\begin{aligned}
	&\frac{\partial P}{\partial x}=-\rho_1\left[\frac{Du}{Dt}-fv\right],\\
	&\frac{\partial P}{\partial y}=-\rho_1\left[\frac{Dv}{Dt}+fu\right],\\
	&\frac{\partial P}{\partial z}=-\rho_1\left[\frac{Dw}{Dt}+g\right],\\
\end{aligned}\right.
\end{equation}
lead to
\begin{equation}\label{press grad}
\left\{\begin{aligned}
	&\frac{\partial P}{\partial x}=-\rho_1\, e^{-ms}\sin\tau \left[k^2\csf^2b+k\csf d f\right],\\
	&\frac{\partial P}{\partial y}=-\rho_1\, e^{-ms}\cos\tau \left[k^2\csf^2d+k\csf b f\right],\\
	&\frac{\partial P}{\partial z}=-\rho_1\, e^{-ms}\left[k^2\csf^2a\, e^{-ms}\cos\tau+g\right].\\
\end{aligned}\right.
\end{equation}
Given \eqref{press grad}, the pressure gradient with respect to the Lagrangian labeling variables 
is given by
\begin{equation}
\begin{pmatrix}
	\frac{\partial P}{\partial q}\\
	\frac{\partial P}{\partial r}\\
	\frac{\partial P}{\partial s}
\end{pmatrix}= \begin{pmatrix}
	\frac{\partial x}{\partial q} & \frac{\partial y}{\partial q} & \frac{\partial z}{\partial q} \\
	\frac{\partial x}{\partial r} & \frac{\partial y}{\partial r} & \frac{\partial z}{\partial r}	\\
	\frac{\partial x}{\partial s} & \frac{\partial y}{\partial s} & \frac{\partial z}{\partial s}
\end{pmatrix}\begin{pmatrix}
	\frac{\partial P}{\partial x}\\
	\frac{\partial P}{\partial y}\\
	\frac{\partial P}{\partial z}\\
\end{pmatrix},
\end{equation}
providing
\begin{equation}\label{222324}
\left\{\begin{aligned}		
	&\frac{\partial P}{\partial q}=-\rho_1 \left\{k^3\csf^2(a^2+d^2-b^2) e^{-2ms}\cos\tau \sin\tau\right.\\
    &\qquad\qquad \left.+(bk^2\csf^2+dkf\csf+gak) e^{-ms} \sin\tau \right\},\\	
	&\frac{\partial P}{\partial r}=-\rho_1 \left\{k\csf(k\csf d + fb) e^{-ms} \cos\tau \right\},\\
	&\frac{\partial P}{\partial s}=-\rho_1 \left\{mk^2\csf^2(a^2+d^2-b^2) e^{-2ms}\cos^2\tau +g \right.+\\
	&\qquad\qquad\left.+(ak^2\csf^2+gam) e^{-ms} \cos\tau +mk\csf(k\csf b^2+fbd)e^{-2ms}\right\}.  
\end{aligned}\right.
\end{equation}
In order to find other two relations for $a, b,d, k, m, \csf,f$, we require the pressure to have continuous second-order partial derivatives, giving
\begin{align}
&  \frac{\partial^2 P}{\partial q\partial r}=\frac{\partial^2 P}{\partial r\partial q}\Rightarrow k\csf d+fb=0,\label{12}\\
& \frac{\partial^2 P}{\partial q\partial s}=\frac{\partial^2 P}{\partial s\partial q}\Rightarrow mk\csf^2b+m\csf df=k^2\csf^2 a .\label{13}
\end{align}
Moreover, as a consequence of \eqref{12} we have that
\begin{equation}\label{P_r}
\frac{\partial P}{\partial r}= 0.
\end{equation}
Therefore, for every constant $P_1$, the gradient of 
\begin{equation}\label{25a}
\begin{aligned}
	P(q-\csf t, r,s )&=P_1-\rho_1 gs +\frac{\rho_1 }{2}(b^2k^2\csf^2+fbdk\csf)e^{-2ms}+\\
    &\hspace{3cm} +\rho_1 (bk\csf^2+df\csf+ga)e^{-ms}\cos\tau+\\
    & \hspace{3cm}+\frac{\rho_1 }{2}k^2\csf^2(a^2+d^2-b^2)e^{-2ms}\cos^2\tau  
\end{aligned}
\end{equation}
with respect to the the labeling variables $(q, s, r)$ gives the right-hand side of \eqref{222324}.\\
Let us rewrite the relations \eqref{9Mc}, \eqref{12} and \eqref{13} as
\begin{align}
&  b=\frac{ma}{k},\label{11bis}\\
&  d=-\frac{fma}{k^2\csf},\label{12bis}\\
&   m^2=\frac{k^4\csf^2}{k^2\csf^2-f^2},\label{13bis}
\end{align}
Observe that \eqref{11bis}, \eqref{12bis} and \eqref{13bis} give also the relation
\begin{equation}\label{adb}
    a^2+d^2=b^2,
\end{equation}
and \eqref{25a} reduces to
\begin{equation}\label{25}
\begin{aligned}
	P(q-\csf t, r,s )=P_1-\rho_1 gs &+\frac{\rho_1 }{2}(b^2k^2\csf^2+fbdk\csf)e^{-2ms}+\\
    &\ +\rho_1 (bk\csf^2+df\csf+ga)e^{-ms}\cos\tau.
\end{aligned}
\end{equation}

\subsection{The layer $\A(t)$ above the halocline }
The layer above the halocline is bounded below by the halocline upper surface $\eta_1$ and above by $\eta_0$. Its internal wave motion is due to the oscillations of the halocline, and---as physical measurements highlights the presence of a mean current of about $0.1\, \mathrm{ms^{-1}}$ in the direction of the Transpolar Drift Current \cite{Guthrie}---we include such feature by assuming a uniform horizontal current in the $x$-direction (in our coordinate system aligned with the TDC). \\
As for the previous layer, the positions of the fluid particles at time $t$ in terms of the labelling variables $(q,\ r,\ s)$  are given by
\begin{equation}\label{LagA}
\left\{	\begin{aligned}
	&x=q-be^{-ms}\sin(k(q-\csf t))-\csf_0 t,\\
	&y=r-de^{-ms}\cos(k(q-\csf t)),\\
	&z=-d_0+s-ae^{-ms}\cos(k(q-\csf t)).
\end{aligned}\right.
\end{equation}
Again, the constant $k=\frac{2\pi}{L}>0$ is the wave number corresponding to the wavelength $L$,  $m>0$  and $a>0$, while the labeling variables are  now chosen so that
\begin{equation}
(q,r,s) \in[-\mathrm{q}_0,\mathrm{q}_0]\times[-\mathrm{r}_0,\mathrm{r}_0]\times [\mathrm{s}_+(r),\mathrm{s}_0(r) ],
\end{equation}
where, as before, $s=\mathrm{s}_+(r)$ represents $\eta_1$, and  $\mathrm{s}_0 $ and $\mathrm{r}_0 $ are defined as previously; moreover $s=\mathrm{s}_0(r)$ represents the upper surface $\eta_0$ of the layer $\A(t)$. The explicit expression for $\eta_0$ is not fundamental and will not be derived. The relevant property of this layer is that, as measurements suggest, it acts as a mean current over the halocline upper surface. \\
The system \eqref{LagA} represents waves travelling in the $x$-direction at a constant speed of propagation $\csf$, in the presence of a constant underlying
current of strength $\csf_0$ ($\csf_0$ is the opposite of the mean Lagrangian velocity in the $x$-direction, c.f. \eqref{c0}).\\
Let us recall the relations between the parameters \eqref{11bis}, \eqref{12bis}, \eqref{13bis} and \eqref{adb}:
\begin{equation}\label{relations}
     b=\frac{ma}{k},\qquad
  d=-\frac{fma}{k^2\csf},\qquad
   m^2=\frac{k^4\csf^2}{k^2\csf^2-f^2}, \qquad a^2+d^2=b^2.
\end{equation}
The particle motion is still trochoidal, with the parameters of the circle \eqref{cerchio} given by
\begin{equation}
    \begin{aligned}
       & \mathbf{C}=(q-\csf_0t,r,s-d_0),   \qquad\mathcal{R}=be^{-ms},\qquad    \theta=\tau=k(q-\csf t),\\
        &\boldsymbol{\nu}=\left(0, \frac{a}{b}, \frac{-d}{b}\right),\qquad
        \boldsymbol{\mu}=\left(0, \frac{-d}{b}, \frac{a}{b}\right),       
    \end{aligned}
\end{equation}
instead of \eqref{cerchioParameters}. Namely, the difference is that the center of the circle $\mathbf{C}$ is moving at constant speed $\csf_0$ along the $x$-axis.\\
The Jacobian of the map \eqref{LagA} is the same as the one of \eqref{Lag}, that is
\begin{equation}\label{18A}
\begin{aligned}
&\left(\frac{\partial(x,y,z)}{\partial(q,r,s)}\right)=\begin{pmatrix}
	\frac{\partial x}{\partial q} & \frac{\partial y}{\partial q} & \frac{\partial z}{\partial q} \\
	\frac{\partial x}{\partial r} & \frac{\partial y}{\partial r} & \frac{\partial z}{\partial r}	\\	
	\frac{\partial x}{\partial s} & \frac{\partial y}{\partial s} & \frac{\partial z}{\partial s}
	
\end{pmatrix}=\\
&\qquad\quad=\begin{pmatrix}
	1-kbe^{-ms}\cos\tau & kde^{-ms}\sin\tau & kae^{-ms}\sin\tau \\
	0 & 1 & 	0\\
	mbe^{-ms}\sin\tau & mde^{-ms}\cos\tau &1+mae^{-ms}\cos\tau
\end{pmatrix},
\end{aligned}
\end{equation}
with
\begin{equation}
\tau=k(q-\csf t),
\end{equation}
and its determinant is equal to
\begin{equation}
J=1-m^2a^2\,e^{-2ms}>0,
\end{equation}
providing, as before, a proof that the the flow described by \eqref{LagA} satisfies the continuity equation \eqref{cont eq}.
The velocity and acceleration of a particle are given by
\begin{equation}\label{20 e 21A}
\left\{\begin{aligned}
	&u=\frac{	Dx}{Dt}=k\csf b\, e^{-ms}\cos\tau-\csf_0,\\
	&v=\frac{	Dy}{Dt}=-k\csf d\, e^{-ms}\sin\tau,\\
	&w=\frac{	Dz}{Dt}=-k\csf a\, e^{-ms}\sin\tau,
\end{aligned}\right.
\quad\text{and} \quad
\left\{\begin{aligned}
	&\frac{	Du}{Dt}=k^2\csf^2 b\, e^{-ms}\sin\tau,\\
	&\frac{	Dv}{Dt}=k^2\csf^2 d\, e^{-ms}\cos\tau,\\
	&\frac{	Dw}{Dt}=k^2\csf^2 a\, e^{-ms}\cos\tau,
\end{aligned}\right.
\end{equation}
and analogous computations as the previous ones give for the pressure (recalling that we are now in the layer with density $\rho_0$ and we made use of \eqref{relations})
\begin{equation}\label{222324A}
\left\{\begin{aligned}		
	&\frac{\partial P}{\partial q}=-\rho_0 \sin\tau+(bk^2\csf^2+dkf\csf+gak-dkf\csf_0) e^{-ms} \sin\tau ,\\	
	&\frac{\partial P}{\partial r}=\rho_0 f \csf_0,\\ 
	&\frac{\partial P}{\partial s}=-\rho_0 \left\{(ak^2\csf^2+gam-mdf\csf_0) e^{-ms} \cos\tau \right.\\
  &\hspace{4cm} \left.+mk\csf(k\csf b^2+fbd)e^{-2ms}+g \right\}.  
\end{aligned}\right.
\end{equation}
For every constant $P_0$, the gradient of 
\begin{equation}\label{25A}
\begin{aligned}
	P(q-\csf t, r,s )=P_0-&\rho_0 gs +\rho_0 f \csf_0 r +\frac{\rho_0 }{2}(b^2k^2\csf^2+fbdk\csf)e^{-2ms}+\\
    &\ +\rho_0 (bk\csf^2+df\csf+ga-df\csf_0)e^{-ms}\cos\tau
\end{aligned}
\end{equation}
with respect to the labelling variables $(q,r,s)$ gives the right-hand side of \eqref{222324A}.\\
The dispersion relation and the expression for the wavenumber $k=\frac{2\pi}{L}$, as well as the existence of $\eta_2$ and $\eta_1$, will be provided by imposing the dynamic boundary condition across the surfaces $\eta_1$ and $\eta_2$.

\subsection{The upper surface of the halocline $\eta_1$}
For every $r\in [-\mathrm{r}_0,\mathrm{r}_0]$, the upper surface of the halocline $\eta_1$ is described by $s=\mathrm{s}_+(r)$, and the continuity of the pressure is imposed by equating \eqref{25} and \eqref{25A} at $s=\mathrm{s}_+(r)$:
\begin{equation}\label{eqP0}
\begin{aligned}
P_0-\rho_0 gs& +\rho_0 f \csf_0 r +\frac{\rho_0 }{2}(b^2k^2\csf^2+fbdk\csf)e^{-2ms}+\\
&+\rho_0 (bk\csf^2+df\csf+ga-df\csf_0)e^{-ms}\cos\tau=\\
&\hspace{2cm}	=P_1-\rho_1 g\mathrm{s}_+ +\frac{\rho_1 }{2}(b^2k^2\csf^2+fbdk\csf)e^{-2m\mathrm{s}_+} +\\
&\hspace{4cm} +\rho_1 (bk\csf^2+df\csf+ga)e^{-m\mathrm{s}_+}\cos\tau.
\end{aligned}
\end{equation}
Observing that, due to \eqref{relations},
\begin{equation}
    k^2\csf^2b^2+fbdk\csf=k^2\csf^2a^2,
\end{equation}
equation \eqref{eqP0} is satisfied if
\begin{equation}\label{a}
\left\{\begin{aligned}
	&        P_0-P_1=\left(\frac{\rho_1-\rho_0}{2}\right) k^2\csf^2a^2e^{-2m\mathrm{s}_+}-(\rho_1-\rho_0)g\mathrm{s}_+ -\rho_0 f \csf_0 r ,\\
	&   \rho_1\left(bk\csf^2+df\csf+ga\right)= \rho_0\left(bk\csf^2+df\csf+ga-df\csf_0\right);
\end{aligned}\right.
\end{equation}
therefore $\eta_1$ is determined by setting $s=\mathrm{s}_+(r)$ at a fixed value of $r\in[-\mathrm{r}_0,\mathrm{r}_0]$, where $\mathrm{s}_+(r)$ is the unique solution of 
\begin{equation}\label{dynamic1}
P_0-P_1=\left(\frac{\rho_1-\rho_0}{2}\right) k^2\csf^2a^2e^{-2m\mathrm{s}_+}-(\rho_1-\rho_0)g\mathrm{s}_+ -\rho_0 f \csf_0 r.
\end{equation}
Let us denote by $A=\left(\frac{\rho_1-\rho_0}{2}\right) k^2\csf^2a^2>0$. For every fixed $r\in[-\mathrm{r}_0,\mathrm{r}_0]$, the function
\begin{equation}
s\longmapsto Ae^{-2m s}-(\rho_1-\rho_0)gs,
\end{equation}
is a strictly decreasing diffeomorphism from $(0,+\infty)$ to $(-\infty,A)$. Consequently, by the implicit function theorem, if 
\begin{equation}
P_0-P_1+\rho_0f\csf_0 r<A,
\end{equation}
for every fixed $r\in[-\mathrm{r}_0,\mathrm{r}_0]$, we can find a unique (smooth) solution $\mathrm{s}_+(r)>0$ of \eqref{dynamic1}. \\
Evaluating \eqref{dynamic1} at $s=\mathrm{s}_+(r)$ and differentiating with respect to $r$ gives
\begin{equation}\label{derivative 1}
\mathrm{s}'_+(r)=\frac{-\rho_0f\csf_0}{2mAe^{-2m \mathrm{s}_+(r)} + (\rho_1-\rho_0)g}.
\end{equation}
Since the denominator in \eqref{derivative 1} is positive, the sign of $\csf_0$ determines whether the upper surface of the halocline is rising or falling with respect to $r$. As the upper surface of the halocline reduces its depth as $r$ increases (namely in the Eurasian Basin), and increases in depth as $r$ decreases (namely in the Amerasian Basin), see \cite{Polyakov2018}, we need to have $\mathrm{s}'_+(r)>0$ in \eqref{derivative 1}, that is to say $\csf_0<0$.\\
Moreover, a careful analysis of the second equation in \eqref{a}
\begin{equation}\label{DynamicDISP}
    \rho_1\left(bk\csf^2+df\csf+ga\right)= \rho_0\left(bk\csf^2+df\csf+ga-df\csf_0\right)
\end{equation}
gives a condition on the sign of $\csf$. More precisely, from \eqref{DynamicDISP}, we can write
\begin{equation}
    (\rho_1-\rho_0)\left(bk\csf^2+df\csf+ga\right)= -\rho_0 df\csf_0 
\end{equation}
that, making use of \eqref{relations}, can be rewritten as 
\begin{equation}\label{condition}
    (\rho_1-\rho_0)\left(g+|\csf|\sqrt{k^2\csf^2-f^2}\right)=\rho_0\frac{f^2|\csf|}{\sqrt{k^2\csf^2-f^2}}\frac{\csf_0}{\csf}.
\end{equation}
The first condition that emerges from \eqref{condition} is 
\begin{equation}\label{C^2}
    \csf^2>\frac{f^2}{k^2},
\end{equation}
and, as the left-hand side of \eqref{condition} is positive and  $\rho_0\frac{f^2|\csf|}{\sqrt{k^2\csf^2-f^2}}>0$, $\csf_0$ and $\csf$ must have the same sign, giving $\csf<0$.
\subsection{The lower surface of the halocline $\eta_2$}
\begin{figure}[h]
\centering
\includegraphics[width=\linewidth]{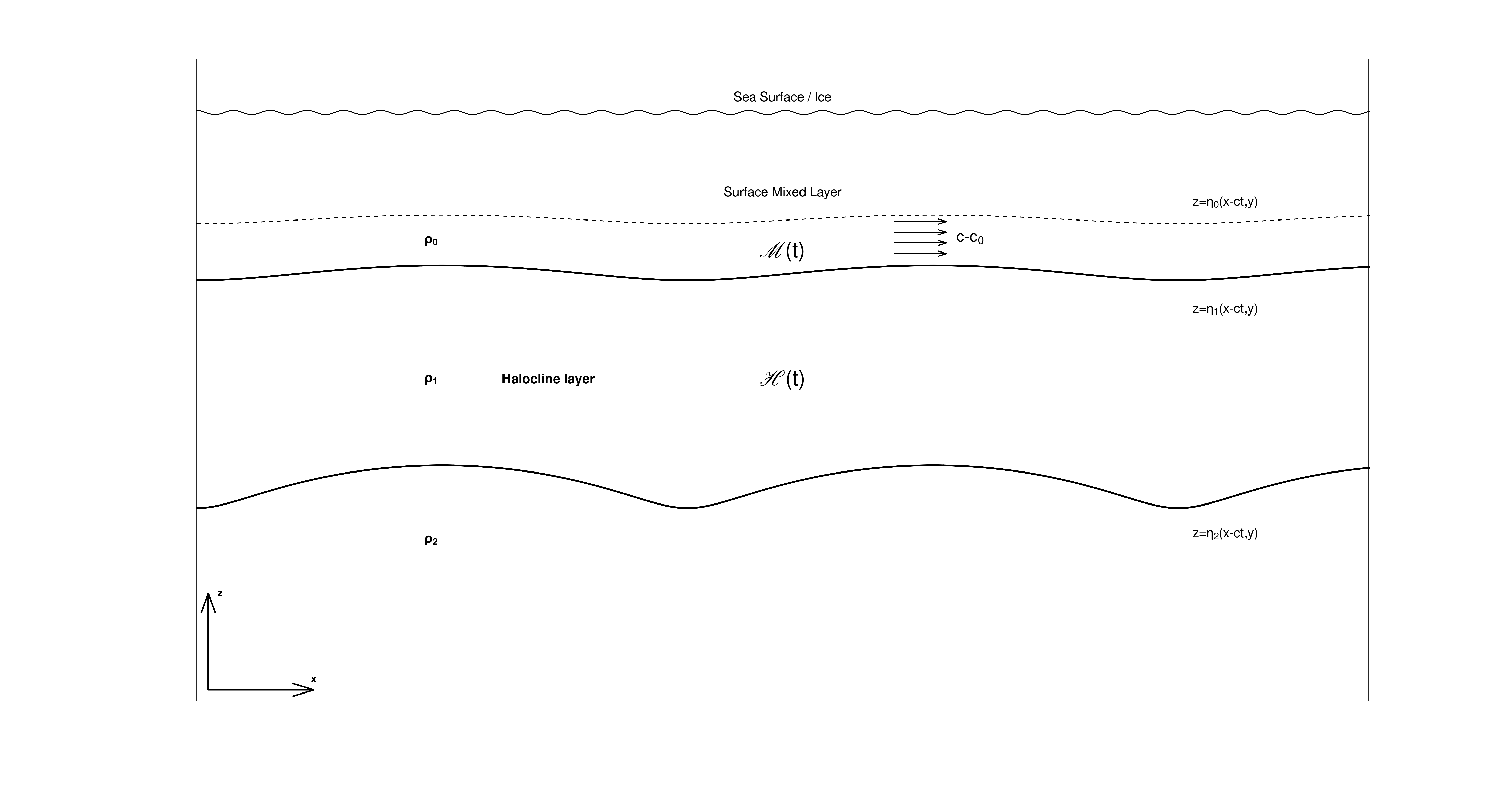}
\caption{Schematic depiction of the flow pattern, at a fixed coordinate $y$}
\label{fig:graphBN}
\end{figure}

For every $r\in [-\mathrm{r}_0,\mathrm{r}_0]$, the lower surface of the halocline $\eta_2$ (also called halocline base) is described by $s=\mathrm{s}_-(r)$.\\
Setting the expression for the pressure in \eqref{25} equal to the one in \eqref{pressure under halocline} at $s=\mathrm{s}_-(r)$, provides the fulfillment of the dynamic boundary condition at $\eta_2$. Therefore we get
\begin{equation}\label{eqP2}
\begin{aligned}
	&P_2+\rho_2 g d_0-\rho_2g\mathrm{s}_- +\rho_2 ga e^{-m \mathrm{s}_-}\cos\tau=\\
	&\hspace{3cm}=P_1-\rho_1 g\mathrm{s}_- +\frac{\rho_1 }{2}(b^2k^2\csf^2+fbdk\csf)e^{-2m\mathrm{s}_-} +\\
    &\hspace{5cm}+\rho_1 (bk\csf^2+df\csf+ga)e^{-m\mathrm{s}_-}\cos\tau,
\end{aligned}
\end{equation}
which is equivalent to
\begin{equation}\label{b}
\left\{\begin{aligned}
	&       P_2-P_1=\frac{1}{2}\rho_1 bk\csf(bk\csf+df)e^{-2m\mathrm{s}_-}+(\rho_2-\rho_1)g\mathrm{s}_- -\rho_2 g d_0,\\
	&      bk\csf^2=\frac{\rho_2}{\rho_1} ga-ga-df\csf.
\end{aligned}\right.
\end{equation}
The halocline base $\eta_2$ is determined by setting $s=\mathrm{s}_-(r)$ at a fixed value of $r\in[-\mathrm{r}_0,\mathrm{r}_0]$, where $\mathrm{s}_-(r)$ is the unique solution of 
\begin{equation}
P_2-P_1=\frac{1}{2}\rho_1 bk\csf(bk\csf+df)e^{-2m s}+(\rho_2-\rho_1)g s-\rho_2 g d_0.
\end{equation}
For every fixed $r\in[-\mathrm{r}_0,\mathrm{r}_0]$, the function
\begin{equation}\label{map}
s\longmapsto \widehat{A} e^{-2m s}+(\rho_2-\rho_1)gs,
\end{equation}
is strictly decreasing if $s<-\frac{1}{2m}\ln\left(\frac{B}{2m\widehat{A}}\right)$, where $B=(\rho_2-\rho_1)g>0$ and  $\widehat{A}=\frac{\rho_1}{2} bk\csf(bk\csf+df)>0$, and strictly increasing if $s>-\frac{1}{2m}\ln\left(\frac{B}{2m\widehat{A}}\right)$, with a minimum at $\tilde{s}=-\frac{1}{2m}\ln\left(\frac{B}{2m \widehat{A} }\right)$.\\
Before proceeding, it is useful to compute the following
\begin{align}\label{valuse}
    \frac{\rho_1-\rho_0}{\rho_0}\approx77.5\cdot10^{-6},\\
        \frac{\rho_2-\rho_1}{\rho_1}\approx443.5\cdot10^{-6},
\end{align}
where me made use of \eqref{densityvariation} and the water parameters presented in the Introduction:
\begin{itemize}
    \item Surface Mixed Layer: water density $\rho_0$, $\mathfrak{T}=-1.5^{\circ}\, \mathrm{C}$, $\mathfrak{S}=34.0\, \mathrm{psu}$;
       \item Halocline: water density $\rho_1$, $\mathfrak{T}=0^{\circ}\, \mathrm{C}$, $\mathfrak{S}=34.2\, \mathrm{psu}$;
       \item Atlantic Water: water density $\rho_2$, $\mathfrak{T}=2^{\circ}\, \mathrm{C}$, $\mathfrak{S}=34.9\, \mathrm{psu}$.
\end{itemize}
We claim that  ${B}>{2m \widehat{A}}$. This will assure that  $s>-\frac{1}{2m}\ln\left(\frac{B}{2m\widehat{A}}\right)$ as, by assumption, $s>0$. As
\begin{equation}
    \frac{B}{2m \widehat{A}}=\frac{1}{ma^2k^2\csf^2}\left(\frac{\rho_2-\rho_1}{\rho_1}\right)g\approx\frac{1.95\cdot10^5}{ma^2},
\end{equation}
using the relation $k^2\csf^2=\frac{f^4 \csf_0^2}{\mathfrak{g}^2} +f^2$, cf. \eqref{redg} and \eqref{dispersionC}, with values $\csf_0\approx 0.1\,\mathrm{ms^{-1}}$, $\mathfrak{g}\approx 8\cdot 10^{-4}\,\mathrm{ms^{-2}}$, $f=2\Omega\approx 1.5\cdot10^{-4}\,\mathrm{s^{-1}}$, and $\left(\frac{\rho_2-\rho_1}{\rho_1}\right)g\approx0.0044\,\mathrm{ms^{-2}}$. As a consequence, $ \frac{B}{2m \widehat{A}}>1$ true if $ma^2<1.95\cdot10^5\mathrm{m^2}$. By \eqref{amax}, $a_{\mathrm{max}}=\frac{1}{m}$, $ma^2<\frac{1}{m}<15$ due to \eqref{m}, it follows that $ \frac{B}{2m \widehat{A}}>1$, and this implies that \eqref{map} is a strictly increasing diffeomorphism from $(0,+\infty)$ to $(A,+\infty)$. Therefore, by the implicit function theorem, there exists a unique (smooth) solution   $\mathrm{s}_-(r)  $. See Figure \ref{fig:graphBN}.\\
\subsection{The dispersion relation}
The last two equations in \eqref{a} and \eqref{b},
namely 
\begin{equation}\label{disp1}
    (\rho_1-\rho_0)\left(bk\csf^2+df\csf+ga\right)= -\rho_0 df\csf_0 
\end{equation}
 and 
 \begin{equation}\label{disp2}
      bk\csf^2+ga+df\csf=\frac{\rho_2}{\rho_1} ga
 \end{equation}
 are independent of $q,r,s$, therefore we can use them to find the dispersion relation of the Pollard waves we are analyzing. Rewriting \eqref{disp1} as
 \begin{equation}\label{disp3}
  bk\csf^2+df\csf+ga= -\frac{\rho_0}{  (\rho_1-\rho_0)} df\csf_0 
\end{equation}
and equaling the left-hand sides of \eqref{disp2} and \eqref{disp3} we arrive at
\begin{equation}
\frac{\rho_2}{\rho_1} ga= -\frac{\rho_0}{  (\rho_1-\rho_0)} df\csf_0 \stackrel{(\ref{relations})}{=} \frac{\rho_0}{  (\rho_1-\rho_0)} \frac{f^2ma}{k^2}\frac{\csf_0}{\csf}\,.
\end{equation}
Defining the reduced gravity
\begin{equation}\label{redg}
\mathfrak{g}:=\left(\frac{\rho_1-\rho_0}{\rho_0}\right) \frac{\rho_2}{\rho_1}g,
\end{equation}
we get 
\begin{equation}\label{disp4}
\mathfrak{g}=\frac{f^2m}{k^2}\frac{\csf_0}{\csf}.
\end{equation} 
Taking the square of both sides of \eqref{disp4} and making use of \eqref{relations} we obtain the dispersion relation 
\begin{equation}\label{dispersionC}
\csf^2= \frac{f^2}{k^2}+\frac{f^4\csf_0^2}{\mathfrak{g}^2k^2}=\frac{f^2}{k^2}\left(1+\frac{f^2\csf_0^2}{\mathfrak{g}^2}\right).
\end{equation} 
Observe that \eqref{dispersionC} satisfies the condition \eqref{C^2}, and as we ask for $\csf<0$, we have that
\begin{equation}
\csf=-\sqrt{\frac{f^2}{k^2}+\frac{f^4\csf_0^2}{\mathfrak{g}^2k^2}}.    
\end{equation}
Since $\mathfrak{g}\approx 8\cdot10^{-4}\,\mathrm{ms^{-2}}$ and $f=2\Omega\approx 1.5\cdot10^{-4}\,\mathrm{s^{-1}}$ we have that $\frac{f^2}{\mathfrak{g}}\approx\frac{1}{30}\,\mathrm{s^2m^{-2}}$.
Moreover, in the ocean, typical fluid velocities rarely exceed $0.1\,\mathrm{ms^{-1}}$, therefore, even setting $\csf_0=1\,\mathrm{ms^{-1}}$ gives $\frac{f^2\csf_0^2}{\mathfrak{g}^2}\approx\frac{1}{30}$, implying that $1
+\frac{f^2\csf_0^2}{\mathfrak{g}^2}\approx1$. As a consequence, 
\begin{equation}
\csf^2\approx \frac{f^2}{k^2},
\end{equation}
giving the modulus of the period of the wave $T=\frac{L}{\csf}$ is approximately $\frac{2\pi}{f}=T_i$, where $T_i$ is the inertial period of the Earth, so the wave motion describing the halocline surfaces is essentially inertial.

\section{Other properties of the flow in the halocline layer $\H(t)$ and in the layer $\A(t)$.}
\subsection{Vorticity}
The bottom layer, being hydrostatic, is irrotational, whereas in the halocline layer $\H(t)$ and in the layer $\A(t)$ above it the vorticity vector
\begin{equation}
\omega=\left(\frac{\partial w}{\partial y}-\frac{\partial v}{\partial z}, \frac{\partial u}{\partial z}-\frac{\partial w}{\partial x}, \frac{\partial v}{\partial x}-\frac{\partial u}{\partial y}\right) \end{equation}
is given by
\begin{equation}\label{vorticity}
\begin{aligned}
	& \omega^T=\frac{1}{1-m^2a^2e^{-2ms}}
	\begin{pmatrix}
	    \frac{m^2 a f}{k}e^{-ms}\sin\tau\\
        \csf a(k^2-m^2)e^{-ms}\cos\tau+\csf m a^2(m^2+k^2)e^{-2ms}\\ fma(\cos\tau+mae^{-ms})e^{-ms}
	\end{pmatrix},
\end{aligned}\end{equation}
due to the relations
\begin{equation}
\begin{aligned}
   & \left(\frac{\partial(q,r,s)}{\partial(x,y,z)}\right)\left(\frac{\partial(u,v,w)}{\partial(q,r,s)}\right)=\left(\frac{\partial(u,v,w)}{\partial(x,y,z)}\right),\\
   &\left(\frac{\partial(q,r,s)}{\partial(x,y,z)}\right)=\left(\frac{\partial(x,y,z)}{\partial(q,r,s)}\right)^{-1},
\end{aligned}
\end{equation}
with
\begin{equation}\label{18inverse}
\begin{aligned}
&  \left(\frac{\partial(q,r,s)}{\partial(x,y,z)}\right)=\begin{pmatrix}
	\frac{\partial q}{\partial x} & \frac{\partial r}{\partial x} & \frac{\partial s}{\partial x} \\
	\frac{\partial q}{\partial y} & \frac{\partial r}{\partial y} & \frac{\partial s}{\partial y} \\
	\frac{\partial q}{\partial z} & \frac{\partial r}{\partial z} & \frac{\partial s}{\partial z} 		
\end{pmatrix}=\\
&\quad\quad=
\begin{pmatrix}
	\frac{1+kbe^{-ms}\cos\tau}{J} & -\frac{kde^{-ms}\sin\tau}{J} & -\frac{kae^{-ms}\sin\tau}{J} \\
	0 & 1 & 	0\\
	-\frac{mbe^{-ms}\sin\tau}{J} & -\frac{mde^{-ms}\cos\tau-m^2ad\, e^{-2ms}}{J} &\frac{1-mae^{-ms}\cos\tau}{J}
\end{pmatrix},  
\end{aligned}
\end{equation}
where $J=1-m^2a^2e^{-2ms}$ is the determinant given in \eqref{jacobian2}, and 
\begin{equation}\label{velocity derivatives}
\begin{aligned}
	&\frac{\p u}{\p q}=-kam\csf\, e^{-ms}\sin\tau,\\
	&\frac{\p v}{\p q}=fma\, e^{-ms}\cos\tau,\\
	&\frac{\p w}{\p q}=-k^2\csf a\, e^{-ms}\cos\tau,
\end{aligned} \qquad
\begin{aligned}
	&\frac{\p u}{\p r}=0,\\
	&\frac{\p v}{\p r}=0,\\
	&\frac{\p w}{\p r}=0,
\end{aligned}
\qquad\begin{aligned}	
&\frac{\p u}{\p s}=-am^2\csf\, e^{-ms}\cos\tau,\\
	&\frac{\p v}{\p s}=-f\frac{am^2}{k}\, e^{-ms}\sin\tau,\\
	&\frac{\p w}{\p s}=km\csf a\, e^{-ms}\sin\tau,
\end{aligned}
\end{equation}
Observe that the expressions \eqref{18inverse} and \eqref{velocity derivatives} are the same both in $\H(t)$ and $\A(t)$, therefore the flow has the same vorticity, equal to \eqref{vorticity}, in the halocline layer and in the layer above it. It is evident that the flow is fully three-dimensional there.\\
Observe that, from \eqref{velocity derivatives}, the velocity field $\bu$ does not depend on the variable $r$, therefore, as 
\begin{equation}
\frac{\partial \bu}{\partial y}= \frac{\partial \bu}{\partial q} \frac{\partial q}{\partial y}+ \frac{\partial \bu}{\partial r} \frac{\partial r}{\partial y}+ \frac{\partial \bu}{\partial s} \frac{\partial s}{\partial y},
\end{equation}
making use of \eqref{18inverse}, it follows that $\bu$ does not depend on the coordinate $y$, namely
\begin{equation}\label{indep Y}
\frac{\partial u}{\partial y}=      \frac{\partial v}{\partial y}=      \frac{\partial w}{\partial y}=0,
\end{equation}
in both the layers $\H(t)$ and $\A(t)$. Moreover, 
\begin{equation}\label{P_y}
\frac{\partial P}{\partial y}= \frac{\partial P}{\partial q} \frac{\partial q}{\partial y}+ \frac{\partial P}{\partial r} \frac{\partial r}{\partial y}+ \frac{\partial P}{\partial s} \frac{\partial s}{\partial y}=\frac{\partial P}{\partial r},
\end{equation}
hence, given \eqref{P_r}, the pressure in independent of $y$ in the halocline layer, while $\frac{\partial P}{\partial y}=\rho_0 f \csf_0$ in the layer above the halocline.
\subsection{Mean flow properties in $\H(t)$}
Following \cite{K2018}, we examine the mean velocities, the Stokes drift, and the mass flux for the internal water waves in the halocline layer. The mean Eulerian velocity is the mean velocity of the fluid at a fixed point, while the mean Lagrangian velocity is the mean velocity following a selected fluid particle. The Stokes drift $\mathbf{U}^S$ is the difference between the mean Lagrangian $\langle \bu \rangle_L$ and the mean Eulerian velocity $\langle \bu \rangle_E$,
\begin{equation}\label{stokesdrift}
\mathrm{U}^S=\langle \bu \rangle_L-\langle \bu \rangle_E,
\end{equation}
and finally we recall that, as noted by Stokes (see \cite{Stokes}), the mass transport is function of the mean Lagrangian velocity rather than the mean Eulerian one.\\
From the expression of the velocities in \eqref{20 e 21}, we can calculate the mean Lagrangian velocities by averaging over a period $T=\frac{L}{\csf}$, obtaining

\begin{equation}\label{mean lag vel}
\begin{aligned}
	&\langle u\rangle_L=\frac{1}{T}\int_0^T k\csf b\, e^{-ms}\cos\tau\, dt=0,\\
	&\langle v\rangle_L=-\frac{1}{T}\int_0^T k\csf d\, e^{-ms}\sin\tau\, dt=0,\\
	&\langle w\rangle_L=-\frac{1}{T}\int_0^T k\csf a\, e^{-ms}\sin\tau\, dt=0.
\end{aligned}
\end{equation}
In order to obtain the mean Eulerian velocity, we need to compute the average of the wave velocity over the period at any fixed depth. As a fixed depth $z_0$ can be written, due to \eqref{Lag}, as
\begin{equation}\label{z_0}
z_0=-d_0+s-ae^{-ms}\cos\tau,
\end{equation}
we can write the functional dependence $s=S(z_0,\tau)$, where $W$ is the Lambert $W$-function, and $S=z_0-W(am\cos\tau e^{mz_0})$ (see \cite{KS}). Taking the derivative with respect to $q$ of \eqref{z_0} with $s=S(z_0,\tau)$, we get
\begin{equation}\label{Sq}
\frac{\partial S}{\partial q}=-\frac{kae^{-ms}\sin\tau}{1+mae^{-ms}\cos\tau}.
\end{equation}
In order to obtain the mean Eulerian $u$-velocity, we write, adding and subtracting $\csf$:
\begin{equation}\label{eulerianU1}
\begin{aligned}
	\langle u\rangle_E(z_0)&=\frac{1}{T}\int_0^T \csf+u(x-\csf t, y, z_0)\, dt-\csf=\\
  &=  \frac{1}{L}\int_0^L \csf+u(x-\csf t, y, z_0)\, dx-\csf=\\
	&=\frac{1}{L}\int_0^L \left[\csf+u(q-\csf t,r, S(z_0,\tau))\right]\frac{\partial x}{\partial q}\, dq-\csf= \\
	&=\frac{1}{L}\int_0^L \csf (1+ ma\, e^{-mS}\cos\tau)\frac{1- m^2a^2\, e^{-2mS}}{1+mae^{-mS}\cos\tau}\, dq-\csf=\\
	&=\csf-\frac{m^2 a^2 \csf}{L}\int_0^L e^{-2mS}\, dq-\csf=-\frac{m^2 a^2 \csf}{L}\int_0^L e^{-2mS}\, dq,
\end{aligned}    
\end{equation} 
so the mean Eulerian $u$-velocity is opposite to the direction of propagation of the Pollard waves (parallel to the Transpolar Drift Current), whereas the other two Eulerian velocities are obtained by direct computations:
\begin{equation}
\begin{aligned}
	\langle v\rangle_E(z_0)&=\frac{1}{T}\int_0^T v(x-\csf t, y, z_0)\, dt=\frac{1}{L}\int_0^L v(x-\csf t, y, z_0)\, dx\\
	&=\frac{1}{L}\int_0^L v(q-\csf t,r, S(z_0,\tau))\frac{\partial x}{\partial q}\, dq= \\
	&=\frac{1}{L}\int_0^L (-k\csf d\, e^{-mS}\sin\tau)\frac{1- m^2a^2\, e^{-2mS}}{1+mae^{-mS}\cos\tau}\, dq=\\
	&=\frac{fma}{kL}\int_0^L  (e^{-mS}\sin\tau)\frac{1- m^2a^2\, e^{-2mS}}{1+mae^{-mS}\cos\tau}\, dq,
\end{aligned}    
\end{equation}

\begin{equation}
\begin{aligned}
	\langle w\rangle_E(z_0)&=\frac{1}{T}\int_0^T w(x-\csf t, y, z_0)\, dt=\frac{1}{L}\int_0^L w(x-\csf t, y, z_0)\, dx\\
	&=\frac{1}{L}\int_0^L w(q-\csf t,r, S(z_0,\tau))\frac{\partial x}{\partial q}\, dq= \\
	&=\frac{1}{L}\int_0^L (-k\csf a\, e^{-mS}\sin\tau)\frac{1- m^2a^2\, e^{-2mS}}{1+mae^{-mS}\cos\tau}\, dq=\\
	&=-\frac{k\csf a}{L}\int_0^L (e^{-mS}\sin\tau)\frac{1- m^2a^2\, e^{-2mS}}{1+mae^{-mS}\cos\tau}\, dq\\
\end{aligned}    
\end{equation}
where we have used \eqref{Lag}, \eqref{9Mc}, \eqref{12bis} and the fact that
\begin{equation}
\frac{\partial x}{\partial q}=1+bm\frac{\partial S}{\partial q}e^{-mS}\sin\tau-kb\,e^{-mS}\cos\tau=\frac{1- m^2a^2\, e^{-2mS}}{1+mae^{-mS}\cos\tau},
\end{equation}
whit the last equality coming from \eqref{Sq}. The components of the Stokes drift, defined in \eqref{stokesdrift}, are therefore given by
\begin{equation}
\begin{aligned}
	U^S&=\frac{m^2 a^2 \csf}{L}\int_0^L e^{-2mS}\, dq,\\
	V^S&=-\frac{fma}{kL}\int_0^L  (e^{-mS}\sin\tau)\frac{1- m^2a^2\, e^{-2mS}}{1+mae^{-mS}\cos\tau}\, dq,\\
	W^S&=\frac{k\csf a}{L}\int_0^L (e^{-mS}\sin\tau)\frac{1- m^2a^2\, e^{-2mS}}{1+mae^{-mS}\cos\tau}\,dq\,.
\end{aligned}    
\end{equation}

Finally, we compute the mass fluxes. As the motion of the water particles in the halocline in three-dimensional, we consider the flux through three orthogonal planes.\\
The mass flux through a plane $\Sigma$ is defined as
\begin{equation}
    \mathcal{M}=\int_{\Sigma} \rho \bu\cdot\boldsymbol{\mathfrak{n}}\, d\Sigma,
\end{equation}
where $\boldsymbol{\mathfrak{n}}$ in the normal vector to the surface $\Sigma$. As we are considering the halocline layer, we set constant density $\rho=\rho_1$.\\
We begin with the mass  flux in the $x$-direction. Let us fix at $x=x_0$ the plane $\Sigma^x=\left[\eta_2,\eta_1\right]\times \left[y_1,y_2\right]$, with $y_1<y_2$. The mass flux in the $x$-direction is given by
\begin{equation}\label{Mx}
\begin{aligned}
     \mathcal{M}^x&=\rho_1 \iint_{\left[\eta_2,\eta_1\right]\times \left[y_1,y_2\right]}  u(x_0-ct,y,z)\, dz\, dy\\
     &=\rho_1 \int_{\mathrm{s_-}}^{\mathrm{s_+}}\int_{-\mathrm{r_0}}^{\mathrm{r_0}}u\det\begin{pmatrix}
        \partial_r y & \partial_s y\\
        \partial_r z & \partial_s z
    \end{pmatrix}dr\, ds.
\end{aligned}
\end{equation}
Having fixed $x=x_0$ implies a functional relation \begin{equation}
    q=\beta(x_0, s,t).
\end{equation}
Taking the $s$-derivative of 
\begin{equation}\label{beta}
    x_0=q-be^{-ms}\sin\tau=\beta(x_0, s,t)-be^{-ms}\sin\tau,
\end{equation}
gives 
\begin{equation}
    \frac{\partial\beta}{\partial s}=-\frac{bme^{-ms}\sin\tau}{1-ma^{-ms}\cos\tau}.
\end{equation}
From \eqref{18}, we have
\begin{equation}
      \frac{\partial y}{\partial r}=1, \quad \frac{\partial z}{\partial r}=0, 
\end{equation}
and, due to the relations
\begin{equation}\label{MX}
    \left.\begin{aligned}
        z=- d_0+s-ae^{-ms}\cos\tau\\
        q=\beta(x_0,s,t)\\
        \frac{\partial\beta}{\partial s}=-\frac{bme^{-ms}\sin\tau}{1-ma^{-ms}\cos\tau}
    \end{aligned}\right\}\Longrightarrow\frac{\partial z}{\partial s}=\frac{1-m^2a^2e^{-2ms}}{1-mae^{-ms}\cos\tau}.
\end{equation}
Therefore, from \eqref{Mx}, the mass flux in the $x$-direction is
\begin{equation}\label{MXX}
    \mathcal{M}^x=2\rho_1 \mathrm{r_0}\csf m a  \int_{\mathrm{s_-}}^{\mathrm{s_+}}e^{-ms}\cos\tau \frac{1-m^2a^2e^{-2ms}}{1-mae^{-ms}\cos\tau}\, ds.
\end{equation}
As $\langle u \rangle_L=0$, we expect that the mass flux \eqref{MX} is zero, when averaged over a wave period $T$. This assertion is in fact true: by differentiating with respect to $t$ \eqref{beta} one gets
\begin{equation}
    \frac{\partial\beta}{\partial t}=-\frac{\csf m a e^{-ms}\cos\tau}{1-ma^{-ms}\cos\tau}
\end{equation}
for the $T$-periodic function $t\mapsto\beta(x_0,s,t)$, hence \eqref{MXX} reads as
\begin{equation}
    \mathcal{M}^x=2\rho_1 \mathrm{r_0}  \int_{\mathrm{s_-}}^{\mathrm{s_+}} \left(m^2a^2e^{-2ms}-1\right)  \frac{\partial\beta}{\partial t}\, ds,
\end{equation}
implying \begin{equation}
    \int_{0}^{T}  \mathcal{M}^x\, dt=0.
\end{equation}

Let us now compute the mass flux in the $y$-direction by fixing at $y=y_0$ the plane $\Sigma^y=\left[\eta_2,\eta_1\right]\times \left[0,L\right]$, where $L=\frac{2\pi}{k}$ is the wave length. The mass flux in the $y$-direction is given by
\begin{equation}\label{My}
\begin{aligned}
     \mathcal{M}^y&=\rho_1 \iint_{\left[\eta_2,\eta_1\right]\times \left[0,L\right]}  v(x-ct,y_0,z)\, dz\, dx\\
     &=\rho_1 \int_{\mathrm{s_-}}^{\mathrm{s_+}}\int_{0}^{L}v\det\begin{pmatrix}
        \partial_q x & \partial_s x\\
        \partial_q z & \partial_s z
    \end{pmatrix}dq\, ds.
\end{aligned}
\end{equation}
As the variables $x$ and $z$ are independent of $r$ (see  \eqref{Lag} and \eqref{18}), there is no need to write a functional relation for fixed  $y=y_0$.
Consequently, the mass flux in the $y$-direction is 
\begin{equation}\label{MYY}
    \mathcal{M}^y=\rho_1 \frac{fma}{k} \int_{\mathrm{s_-}}^{\mathrm{s_+}}\int_0^L e^{-ms}\sin\tau \left(1-m^2a^2e^{-2ms}\right)\,dq ds,
\end{equation}
where we also used the relation \eqref{12bis}. It is immediate to see that
\begin{equation}
    \int_{0}^{T}  \mathcal{M}^y\, dt=0,
\end{equation}
as expected, since $\langle v \rangle_L=0$.\\
Finally, for the mass flux in the $z$-direction, let us fix at $z=z_0$ the plane $\Sigma^z=[0,L]\times[y_1,y_2]$, where $z_0$ is chosen between the crest $\mathrm{z}_-$ of the halocline lower surface $\eta_2$ and the trough $\mathrm{z}_+$ of the halocline upper surface $\eta_1$, that is $\mathrm{z}_-<z_0<\mathrm{z}_+$.  We write the functional relation 
\begin{equation}
    s=\xi(z_0, q,t).
\end{equation}
Writing, according to \eqref{Lag}
\begin{equation}
    z_0=s - a e^{-ms}\cos\tau,
\end{equation}
we get
\begin{align}
    \frac{\partial\xi}{\partial q}&=-\frac{ka e^{-m\xi}\sin\tau}{1+ma e^{-m\xi}\cos\tau},\label{xi-q}\\
     \frac{\partial\xi}{\partial t}&=\frac{k\csf a e^{-m\xi}\sin\tau}{1+ma e^{-m\xi}\cos\tau}.\label{xi-t}
\end{align}
Equation \eqref{xi-q} gives
\begin{equation}
    \frac{\partial x}{\partial q}=\frac{1-m^2a^2 e^{-2m\xi}}{1+ma e^{-m\xi}\cos\tau},
\end{equation}
and, since from \eqref{18} we have
\begin{equation}
      \frac{\partial x}{\partial r}=0, \quad   \frac{\partial y}{\partial r}=1,
\end{equation}
the mass  flux in the $z$-direction, defined as
\begin{equation}\label{Mz}
\begin{aligned}
     \mathcal{M}^z&=\rho_1 \iint_{ \left[0,L\right]\times[y_1,y_2]}  w(x-ct,y,z_0)\, dz\, dx\\
     &=\rho_1 \int_{0}^{L}\int_{-\mathrm{r_0}}^{\mathrm{r_0}}w\det\begin{pmatrix}
        \partial_q x & \partial_r x\\
        \partial_q y & \partial_r y
    \end{pmatrix}dq\, dr,
\end{aligned}
\end{equation}
is given by
\begin{equation}\label{MZZ}
    \mathcal{M}^z=-2\rho_1 \mathrm{r_0} \int_{0}^{L}k\csf a e^{-m\xi}\sin\tau \frac{1-m^2a^2e^{-2m\xi}}{1+mae^{-m\xi}\cos\tau}\, dq.
\end{equation}
Furthermore, as
\begin{equation}\label{MZt}
    \mathcal{M}^z=-2\rho_1 \mathrm{r_0} \int_{0}^{L}\left(1-m^2a^2e^{-2m\xi}\right)\frac{\partial \xi}{\partial t}\, dq,
\end{equation}
due to \eqref{xi-t}, and the function $t\mapsto\xi(z_0,q,t)$ being $T$-periodic, we get

\begin{equation}
    \int_0^T\mathcal{M}^z\, dt=0,
\end{equation}
as could be inferred by $\langle w\rangle_L=0.$\\
In conlusion, this prove that the Pollard internal wave has no net wave transport over a wave period.

\subsection{Mean flow properties in $\A(t)$}
As in the layer above the halocline the only difference in the velocity field, with respect to the one in the halocline layer, is in $u$, being equal to
\begin{equation}\label{uA}
    u=k\csf b\, e^{-ms}\cos\tau-\csf_0
\end{equation}
in place of $ u=k\csf b\, e^{-ms}\cos\tau$, we analyze only the mean flow properties involving $u$, namely $\langle u \rangle_L,\ \langle u \rangle_E,\ \mathrm{U}^S $ 
and $\mathcal{M}^x$.
From the expression \eqref{uA}, we can calculate, as before 
\begin{equation}\label{c0}
	\langle u\rangle_L=\frac{1}{T}\int_0^T (k\csf b\, e^{-ms}\cos\tau-\csf_0)\, dt=-\csf_0,
\end{equation}
where $T=\frac{L}{\csf}$ is a wave period. To obtain the mean Eulerian velocity, we need to compute the average of the wave velocity over the period at any fixed depth. As a fixed depth $z_0$ can still be written  as
\begin{equation}\label{z_0A}
z_0=-d_0+s-ae^{-ms}\cos\tau,
\end{equation}
we can write the same functional dependence $s=S(z_0,\tau)$ as before to get
\begin{equation}\label{SqA}
\frac{\partial S}{\partial q}=-\frac{kae^{-ms}\sin\tau}{1+mae^{-ms}\cos\tau}.
\end{equation}
For the mean Eulerian $u$-velocity, as before, we add and subtract $\csf$:
\begin{equation}\label{eulerianU2}
\begin{aligned}
	\langle u\rangle_E(z_0)&=\frac{1}{T}\int_0^T \csf+u(x-\csf t, y, z_0)\, dt-\csf=\\
    &=\frac{1}{L}\int_0^L \csf+u(x-\csf t, y, z_0)\, dx-\csf=\\
	&=\frac{1}{L}\int_0^L \left[\csf+u(q-\csf t,r, S(z_0,\tau))\right]\frac{\partial x}{\partial q}\, dq-\csf= \\
	=-\csf&+\frac{1}{L}\int_0^L \csf (1-\frac{\csf_0}{\csf}+ ma\, e^{-mS}\cos\tau)\frac{1- m^2a^2\, e^{-2mS}}{1+mae^{-mS}\cos\tau}\, dq=\\
	&=-\frac{m^2 a^2 \csf}{L}\int_0^L e^{-2mS}\, dq-\frac{\csf_0}{L}\int_0^L \frac{1- m^2a^2\, e^{-2mS}}{1+mae^{-mS}\cos\tau}\, dq
\end{aligned}    
\end{equation} 
and, consequently
\begin{equation}
	U^S=\frac{m^2 a^2 \csf}{L}\int_0^L e^{-2mS}\, dq+\frac{\csf_0}{L}\int_0^L \frac{1- m^2a^2\, e^{-2mS}}{1+mae^{-mS}\cos\tau}\, dq-\csf_0.
\end{equation}
Note that also in this layer the mean Eulerian velocity in the $x$-direction \eqref{eulerianU2} is positive. Finally,  the mass  flux in the $x$-direction is given by
\begin{equation}\label{MxA}
\begin{aligned}
     \mathcal{M}^x&=\rho_0 \iint_{\left[\eta_1,\eta_0\right]\times \left[y_1,y_2\right]}  u(x_0-ct,y,z)\, dz\, dy\\
     &=\rho_1 \int_{\mathrm{s_+}}^{\mathrm{s_0}}\int_{-\mathrm{r_0}}^{\mathrm{r_0}}u\det\begin{pmatrix}
        \partial_r y & \partial_s y\\
        \partial_r z & \partial_s z
    \end{pmatrix}dr\, ds.
\end{aligned}
\end{equation}
Again, having fixed $x=x_0$ implies a functional relation \begin{equation}
    q=\beta_0(x_0, s,t),
\end{equation}
and taking the $s$-derivative of 
\begin{equation}
    x_0=q-be^{-ms}\sin\tau-\csf_0 t=\beta_0(x_0, s,t)-be^{-ms}\sin\tau-\csf_0 t,
\end{equation}
gives 
\begin{equation}
    \frac{\partial\beta_0}{\partial s}=-\frac{bme^{-ms}\sin\tau}{1-ma^{-ms}\cos\tau}.
\end{equation}
As
\begin{equation}
      \frac{\partial y}{\partial r}=1, \qquad \frac{\partial z}{\partial r}=0, \qquad \frac{\partial z}{\partial s}=\frac{1-m^2a^2e^{-2ms}}{1-mae^{-ms}\cos\tau},
\end{equation}
 the mass flux in the $x$-direction is
\begin{equation}\label{MXA}
\begin{aligned}
     \mathcal{M}^x&=2\rho_0 \mathrm{r_0}   \int_{\mathrm{s_+}}^{\mathrm{s_0}}(\csf m ae^{-ms}\cos\tau-\csf_0) \frac{1-m^2a^2e^{-2ms}}{1-mae^{-ms}\cos\tau}\, ds\\
     &\hspace{2cm}=2\rho_0 \mathrm{r_0}   \csf m a\int_{\mathrm{s_+}}^{\mathrm{s_0}}e^{-ms}\cos\tau \frac{1-m^2a^2e^{-2ms}}{1-mae^{-ms}\cos\tau}\, ds-\\
     &\hspace{4.5cm}-2\rho_0 \mathrm{r_0} \csf_0 \int_{\mathrm{s_+}}^{\mathrm{s_0}} \frac{1-m^2a^2e^{-2ms}}{1-mae^{-ms}\cos\tau}\, ds.
\end{aligned}
\end{equation}
Averaging $ \mathcal{M}^x$ over a wave period $T$ we get
\begin{equation}\label{MXTA}
    \frac{1}{T}\int_0^T \mathcal{M}^x\, dt= -  2\frac{\rho_0 \mathrm{r_0} \csf_0}{T}\int_0^T\int_{\mathrm{s_+}}^{\mathrm{s_0}} \frac{1-m^2a^2e^{-2ms}}{1-mae^{-ms}\cos\tau}\, dsdt\not=0,
\end{equation} and this reflects the fact that in the layer above the halocline $\langle u\rangle_L=-\csf_0\not=0$. \\
In \eqref{MXTA} we have used again the fact that the function $t\mapsto\beta_0(x_0,s,t)$ is $T$-periodic and that
\begin{equation}
    \frac{\partial\beta_0}{\partial t}=-\frac{\csf m a e^{-ms}\cos\tau}{1-ma^{-ms}\cos\tau}.
\end{equation}
This shows that, as one could have expected, there is mass transport in the $x$-direction in the layer $\A(t)$.

\section{Comparison with the linearized Lagrangian version of the problem}
Linearizing about the hydrostatic solution $x=q,\ y=r,\ z=s$ the governing equations in Lagrangian form (see \cite{KS} and \cite{Pierson}), namely
\begin{equation}\label{Euler LAG} 
\left\{\begin{aligned}
	&\left(\frac{\partial^2 x} {\partial t^2}  	-f\frac{\partial y}{\partial t} \right) \frac{\partial x}{\partial q}+ \left(\frac{\partial^2 y} {\partial t^2}  	+f\frac{\partial x}{\partial t} \right) \frac{\partial y}{\partial q}+\frac{\partial^2 z} {\partial t^2}   \frac{\partial x}{\partial q}+g \frac{\partial z}{\partial q}=
	-\frac{1}{\rho}
	\frac{\partial P}{\partial q}, \\
	&\left(\frac{\partial^2 x} {\partial t^2}  	-f\frac{\partial y}{\partial t} \right) \frac{\partial x}{\partial r}+ \left(\frac{\partial^2 y} {\partial t^2}  	+f\frac{\partial x}{\partial t} \right) \frac{\partial y}{\partial r}+\frac{\partial^2 z} {\partial t^2}   \frac{\partial x}{\partial r}+g \frac{\partial z}{\partial r}=
	-\frac{1}{\rho}
	\frac{\partial P}{\partial r}, \\
    &\left(\frac{\partial^2 x} {\partial t^2}  	-f\frac{\partial y}{\partial t} \right) \frac{\partial x}{\partial s}+ \left(\frac{\partial^2 y} {\partial t^2}  	+f\frac{\partial x}{\partial t} \right) \frac{\partial y}{\partial s}+\frac{\partial^2 z} {\partial t^2}   \frac{\partial x}{\partial s}+g \frac{\partial z}{\partial q}=
	-\frac{1}{\rho}
	\frac{\partial P}{\partial s}, 
\end{aligned}\right.
\end{equation}
and 
\begin{equation}\label{inc Lag}
    \frac{d}{dt}\left(\det \left(\frac{\partial(x,y,z)}{\partial(q,r,s)}\right)\right)=0,
\end{equation}
which are the Euler equations and the incopressibility condition, respectively, yield to the linearized Euler equations
\begin{equation}\label{Linear LAG} 
\left\{\begin{aligned}
	&\frac{\partial^2 x} {\partial t^2}  	-f\frac{\partial y}{\partial t} + g \frac{\partial z}{\partial q}=
	-\frac{1}{\rho}
	\frac{\partial P}{\partial q}, \\
	&\frac{\partial^2 y} {\partial t^2} + 		f \frac{\partial x}{\partial t} + g \frac{\partial z}{\partial r}=
	-\frac{1}{\rho}
	\frac{\partial P}{\partial r}, \\
	&\frac{\partial^2 z} {\partial t^2}  + g \frac{\partial z}{\partial s}
	=
	-\frac{1}{\rho}
	\frac{\partial P}{\partial s} 	,
\end{aligned}\right.
\end{equation}
and the linearized incompressibility condition
\begin{equation}\label{linear inc}
    \frac{\partial^2 x}{\partial q \partial t}+\frac{\partial^2 y}{\partial r \partial t}+\frac{\partial^2 z}{\partial s \partial t}=0.
\end{equation}
Remarkably, Pollard waves \eqref{Lag} and \eqref{LagA} are solutions also to the system of linearized equations \eqref{Linear LAG} and \eqref{linear inc} in the layers $\H(t)$ and $\A(t)$ respectively. Namely, writing
\begin{equation}\label{Lag LIN}
\left\{	\begin{aligned}
	&x=q-be^{-ms}\sin\tau,\\
	&y=r-de^{-ms}\cos\tau,\\
	&z=-d_0+s-ae^{-ms}\cos\tau,
\end{aligned}\right.  
\end{equation}
with $(q,r,s) \in[-\mathrm{q}_0,\mathrm{q}_0]\times[-\mathrm{r}_0,\mathrm{r}_0]\times [\mathrm{s}_-(r),\mathrm{s}_+(r) ]$ for the halocline, and 
\begin{equation}\label{Lag LIN A}
\left\{	\begin{aligned}
	&x=q-be^{-ms}\sin\tau-\csf_0t,\\
	&y=r-de^{-ms}\cos\tau,\\
	&z=-d_0+s-ae^{-ms}\cos\tau,
\end{aligned}\right. \end{equation}
with  $(q,r,s) \in [-\mathrm{q}_0,\mathrm{q}_0]\times[-\mathrm{r}_0,\mathrm{r}_0]\times [\mathrm{s}_+(r),\mathrm{s}_0(r) ]$
for the layer above it, the linearized incompressibility equation \eqref{linear inc} is equivalent, in both layers, to the condition \eqref{9Mc}, that is
\begin{equation}\label{c1}
    b=\frac{am}{k}.
\end{equation}
Rewriting \eqref{Linear LAG} in $\H(t)$ as
\begin{equation}\label{Linear LAG 2}
\left\{\begin{aligned}
	&
	\frac{\partial P}{\partial q}=-\rho_1\left[\frac{\partial^2 x} {\partial t^2}  	-f\frac{\partial y}{\partial t} + g \frac{\partial z}{\partial q} \right], \\
	&	\frac{\partial P}{\partial r}=-\rho_1\left[\frac{\partial^2 y} {\partial t^2} + 		f \frac{\partial x}{\partial t} + g \frac{\partial z}{\partial r} \right], \\
	& 	\frac{\partial P}{\partial s}=-\rho_1\left[\frac{\partial^2 z} {\partial t^2}  + g \frac{\partial z}{\partial s} \right]	,
\end{aligned}\right.
\end{equation}
gives, for the pressure, the solution
\begin{equation}\label{linearPressure}
    P=\widetilde{P_1}-\rho_1 gs +\rho_1\left[\frac{k^2\csf^2 a}{m}+ga\right]e^{-ms}\cos\tau \qquad\text{in}\ \H(t),
\end{equation}
with the conditions 
\begin{equation}\label{c2}
    k\csf d + fb=0,
\end{equation}
and
\begin{equation}\label{c3}
m^2=\frac{k^4\csf^2}{\csf^2k^2-f^2},
\end{equation}
whereas, rewriting \eqref{Linear LAG} in $\A(t)$ in an analogous manner (i.e. with $\rho_0$ in place of $\rho_1$ and using \eqref{Lag LIN A} instead of \eqref{Lag LIN}) gives
\begin{equation}\label{linearPressure0}
    P=\widetilde{P_0}-\rho_0 gs +\rho_0\left[\frac{k^2\csf^2 a}{m}+ga\right]e^{-ms}\cos\tau +\rho_0f\csf_0 r \qquad\text{in}\ \A(t).
\end{equation}

We remark that the conditions \eqref{c1}, \eqref{c2} and \eqref{c3} are exactly those found in the nonlinear analysis: \eqref{9Mc}, \eqref{12} and \eqref{13}.\\
The continuity of the pressure across the halocline top surface gives the following two equations:
\begin{align}
	&        \widetilde{P_1}-\widetilde{P_0}=(\rho_1-\rho_0)g\mathrm{s}_+ +\rho_0 f \csf r ,\\
	&   \rho_0\left[\frac{k^2\csf^2 a}{m}+ga\right]=\rho_1\left[\frac{k^2\csf^2 a}{m}+ga\right] .\label{problema}
\end{align}
where $s=\mathrm{s}_+$ represent the upper surface of the halocline $\eta_1$.\\
As $\rho_1>\rho_0$, it is evident that \eqref{problema} cannot be true. This shows that the linear version of the problem is not suitable for the analysis we have made, highlighting the importance of the nonlinearity in the governing equations.

\section{Discussion}
We conclude by presenting some qualitative and quantitative considerations for the solution \eqref{Lag} and \eqref{LagA}.\\
Recalling that the particle motion in \eqref{Lag} describes trochoidal orbits, namely the path of a fixed point on a circle of radius $be^{-ms}$ and centred at $(q,r,s-d_0)$ (eventually rolling along the $x$-axis in the the case of \eqref{LagA}), in a plane that is at an angle $\arctan\left(-\frac{d}{a}\right)$ with respect to the vertical axis, due to \eqref{12bis}, \eqref{13bis} and \eqref{dispersionC}, we have that
\begin{equation}
    \arctan\left(-\frac{d}{a}\right)= \arctan\left(\frac{\mathfrak{g}}{f|\csf_0|}\right).
\end{equation}
As $\mathfrak{g}\approx8\cdot10^{-4}\, \mathrm{m s^{-2}}$ and $f\approx 1.5\cdot10^{-4}\, \mathrm{s^{-1}}$, and with $|\csf_0|\approx 0.1\, \mathrm{ms^{-1}}$, we have that $\arctan\left(\frac{\mathfrak{g}}{f|\csf_0|}\right)\approx 89^{\circ}$, and such angle increases as $|\csf_0|$ decreases. Note that, even with the extreme value $|\csf_0|\approx 1\, \mathrm{ms^{-1}}$, this angle is about  $80^{\circ}$.
Namely, the trochoidal orbits described in \eqref{Lag} and \eqref{LagA} are almost horizontal.\\
We point out that, even if the wave motion has a negative wave-speed $\csf$ (i.e. waves move in the direction opposite to $x$-direction), the mean Eulerian velocity in the $x$-direction is in the positive direction. In fact $\langle u\rangle_E$ ia given by (c.f. \eqref{eulerianU1} and \eqref{eulerianU2})
\begin{equation}
    	\langle u\rangle_E=\begin{cases}    	    
    	-\frac{m^2 a^2 \csf}{L}\int_0^L e^{-2mS}\, dq & \text{if}\quad \eta_2\leq z\leq \eta_1,\\
        -\frac{m^2 a^2 \csf}{L}\int_0^L e^{-2mS}\, dq-\frac{\csf_0}{L}\int_0^L \frac{1- m^2a^2\, e^{-2mS}}{1+mae^{-mS}\cos\tau}\, dq    
    	 & \text{if}\quad \eta_1 < z\leq \eta_0,
        \end{cases}
\end{equation}
where 
\begin{equation}
\begin{aligned}    	    
    	&\frac{m^2 a^2 }{L}\int_0^L e^{-2mS}\, dq >0,\\
   &    \frac{1}{L}\int_0^L \frac{1- m^2a^2\, e^{-2mS}}{1+mae^{-mS}\cos\tau}\, dq  >\frac{1}{2}-\frac{m^2 a^2 }{2L}\int_0^L e^{-2mS}\, dq >0,
        \end{aligned}
\end{equation} 
and recalling that $\csf,\, \csf_0<0.$      \\
From the dispersion relation \eqref{dispersionC}
\begin{equation}\label{C,K}
\csf^2= \frac{f^2}{k^2}+\frac{f^4\csf_0^2}{\mathfrak{g}^2k^2}\Longrightarrow k^2\csf^2=f^2\left(1+\frac{f^2\csf_0^2}{\mathfrak{g}^2}\right),
\end{equation} 
where $\mathfrak{g}:=\left(\frac{\rho_1-\rho_0}{\rho_0}\right) \frac{\rho_2}{\rho_1}g
\approx 8\cdot 10^{-4}\, \mathrm{ms^{-2}}$ and $f=2\Omega\approx 1.5\cdot10^{-4}\,\mathrm{s^{-1}}$, due to \eqref{13bis} and \eqref{C,K}, we can write
\begin{equation}\label{m2}
    m^2=\frac{ \mathfrak{g}^2 k^4\csf^2}{f^4\csf_0^2}=\frac{ \mathfrak{g}^2 k^2}{f^2\csf_0^2}\left(1+\frac{f^2\csf_0^2}{\mathfrak{g}^2}\right)\approx\frac{ \mathfrak{g}^2 k^2}{f^2\csf_0^2}.
\end{equation}
Fixing $\csf_0\approx-0.1\, \mathrm{ms^{-1}}$ we have the following values of $m$ as a function of $k$:
\begin{equation}\label{m}
    m\approx\frac{ \mathfrak{g}}{f|\csf_0|}k\approx\begin{cases}
        0.08\, \mathrm{m^{-1}}\qquad \text{for}\quad k=0.0015\, \mathrm{m^{-1}}\quad \text{i.e.}\quad L\approx 4.2\, \mathrm{km},\\
         0.33\, \mathrm{m^{-1}}\qquad \text{for}\quad k=\frac{\pi}{500}\, \mathrm{m^{-1}}\quad \text{i.e.}\quad L= 1\, \mathrm{km},\\
         1\, \mathrm{m^{-1}}\qquad \text{for}\quad k=\frac{3}{160}\, \mathrm{m^{-1}}\quad \text{i.e.}\quad L\approx 335\, \mathrm{m}.\\
    \end{cases}
\end{equation}
The three wavenumbers in \eqref{m} corresponds, respectively, to $\csf\approx-0.1\, \mathrm{ms^{-1}}$, $\csf\approx-0.024\, \mathrm{ms^{-1}}$ and $\csf\approx-0.008\, \mathrm{ms^{-1}}$, whereas a wave-speed of around  $\csf\approx-1\, \mathrm{ms^{-1}}$ correspond to $k\approx 1.5 \cdot 10^{-4}\,  \mathrm{m^{-1}}$ (relative to a wavelength $L\approx 42\,  \mathrm{km}$---a value too high to be physically relevant).\\
From \eqref{Lag} and \eqref{LagA} it is immediate to see that as depth increases, the amplitude of the waves increases and the maximum oscillation of the wave will change according to the formula $a_{\mathrm{max}}=\frac{1}{m}$ the values in \eqref{m}, depending on the wavelength (or wavenumber). As it is evident from \eqref{m}, the maximum amplitude of the wave will be higher for higher wavelengths, and smaller for smaller wavelengths.\\
Moreover, as the amplitude of the waves described in \eqref{Lag} (and \eqref{LagA}) increases when the depth of the halocline reduces, we can infer that in summer---with the decreasing of the depths of the halocline boundaries \cite{Peralta}---the amplitude of the oscillations will decrease, and vice-versa in winter.\\
About the halocline's upper surface, as anticipated, \eqref{derivative 1} shows that the upper surface of the halocline reduces its depth as $r$ increases (namely in the Eurasian Basin), and increases in depth as $r$ decreases (namely in the Amerasian Basin). This result matches the measurements made by oceanographers (see \cite{Polyakov2018}), even if we infer that there should be other physical processes influencing the halocline depth, such as water inflows from other basins, against these variations. In fact, from \eqref{derivative 1} we have that (with $|\csf_0|\approx0.1\, \mathrm{m s^{-1}}$) 
\begin{equation}\label{derivative 2}
\mathrm{s}'_+(r)=\frac{\rho_0f|\csf_0|}{2mAe^{-2m \mathrm{s}_+(r)} + (\rho_1-\rho_0)g}<\left(\frac{ (\rho_1-\rho_0)g}{\rho_0}\right)^{-1}f|\csf_0|\lessapprox 0.019,
\end{equation}
thus, an increase [decrease] in $r$ of kilometres produces a decrease [increase] in the depth of the halocline's upper surface of tens of meters. As depth variations of the halocline are less than a hundred meters in different basins (see \cite{Polyakov2018} or \cite{Rudels}), we deduce that such variations must also be controlled by other mechanisms.\\
One downside of our model is that it does not provide information on the increase/decrease of the halocline base. A better model that would provide such information will be the subject of a subsequent work.\\
We conclude by highlighting two fundamental aspects of our analysis: firstly,  the simple near-inertial dispersion relation obtained in \eqref{dispersionC}, imposed by the two dynamic boundary conditions at the halocline surfaces, in contrast to other works---with only one dynamic boundary condition---having much more complicated dispersion relations, describing two wave modes (one slow, near inertial and one faster) whose formulae were obtained via a perturbative analysis of the dispersion relation, and not in closed form \cite{Constantin2017}, \cite{McCarney2023}, \cite{McCarney2024}. As pointed out by Garret \& Munk (see \cite{GM1} and \cite{GM2}), near-inertial waves are the most energetic ones, but are hard to detect, even without the presence of ice (e.g. using satellite methods \cite{WP1996}).\\
Secondly, our analysis has shown that the linear version of the problem is not sufficient to describe the model considered, and how the nonlinearity in the governing equations---despite making them more complicated--plays a fundamental role in obtaining a relevant solution.
\section*{Funding}
The author is supported by the Austrian Science Fund (FWF) [grant number Z 387-N, grant doi: https://doi.org/10.55776/Z387].

\section*{Acknowledgments}
The author gratefully acknowledges Luigi Roberti for his valuable suggestions and the two anonymous reviewers for their insightful comments and remarks, which helped improve the manuscript.

\end{document}